\documentclass[fleqn,usenatbib]{mnras}

\usepackage[T1]{fontenc}
\usepackage{ae,aecompl}

\usepackage{graphicx}	
\usepackage{amsmath}	
\usepackage{amssymb}

\usepackage{txfonts}	

\newcommand{\btxt}[1]{{#1}}
\newcommand{\ctxt}[1]{{#1}}

\title[Proper motions of bow shock stars]{Proper motions of five OB
  stars with candidate dusty bow shocks in the Carina Nebula}

\author[M. M. Kiminki et al.]{
Megan M. Kiminki,$^{1}$\thanks{E-mail: \href{mailto:mbagley@email.arizona.edu}{mbagley@email.arizona.edu}}
Nathan Smith,$^{1}$
Megan Reiter$^{2}$
and
John Bally$^{3}$
\\
$^{1}$Steward Observatory, University of Arizona, 933 N.
  Cherry Avenue, Tucson, AZ 85721, USA
\\
$^{2}$Department of Astronomy, University of Michigan, 311 West Hall, 1085 S. University Avenue, Ann Arbor, MI 48109, USA
\\
$^{3}$Department of Astrophysical and Planetary Sciences, University of Colorado, UCB 389, Boulder, CO 80309, USA
}

\date{Accepted 2017 March 9. Received 2017 March 7; in original form 2016 November 21.}

\pubyear{2017}

\begin{document}
\label{firstpage}
\pagerange{\pageref{firstpage}--\pageref{lastpage}}
\maketitle

%%%%%%%%%%%%%%%%%%%%%%%%%%%%%%%%%%%%%%%%%%%%%%%%%%%%%%%%%%%%%%%%%%%%%%%
\begin{abstract}
We constrain the proper motions of five OB stars associated with
candidate stellar wind bow shocks in the Carina Nebula using
\emph{HST} ACS imaging over 9--10 year baselines.  These proper
motions allow us to directly compare each star's motion to the
orientation of its \btxt{candidate} bow shock.  Although these stars
are saturated in our imaging, we assess their motion by the shifts
required to minimize residuals in their Airy rings.  The results limit
the direction of each star's motion to sectors less than 90$\degr$
wide.  None of the five stars are moving away from the Carina Nebula's
central clusters as runaway stars would be, confirming that \btxt{a
  candidate} bow shock is not necessarily indicative of a runaway
star.  Two of the five stars are moving tangentially relative to the
orientation of their \btxt{candidate} bow shocks, both of which point
at the OB cluster Trumpler 14.  In these cases, the large-scale flow
of the interstellar medium, powered by feedback from the cluster,
appears to dominate over the motion of the star in producing the
observed \btxt{candidate} bow shock.  The remaining three stars all
have some component of motion toward the central clusters, meaning
that we cannot distinguish whether their \btxt{candidate} bow shocks
are indicators of stellar motion, of the flow of ambient gas, or of
density gradients in their surroundings.  In addition, these stars'
lack of outward motion hints that the distributed massive-star
population in Carina's South Pillars region formed in place, rather
than migrating out from the association's central clusters.
\end{abstract}


%%%%%%%%%%%%%%%%%%%%%%%%%%%%%%%%%%%%%%%%%%%%%%%%%%%%%%%%%%%%%%%%%%%%%%%
\section{Introduction}
\label{sec:intro}

Feedback from massive stars impacts their surroundings on scales
ranging from the shaping of their immediate circumstellar environment
to the reionization of the universe.  Stellar wind bow shocks, falling
on the former end of that scale, provide important information about a
star's history and environment.  Bow shocks are produced when the
relative velocity between a star and the surrounding interstellar
medium (ISM) is supersonic \citep{baranov1971,vanburenmccray1988}.
They typically appear as arc-shaped features in optical line emission
\citep[e.g.,][]{kaper1997,bally2000,brownbomans2005,brownsbergerromani2014}
and/or thermal infrared continuum emission from dust
\citep[e.g.,][]{vanburen1995,noriegacrespo1997,comeronpasquali2007,france2007,gaspar2008,peri2012,winston2012,kobulnicky2016}.
These features mark the sweeping-up of ambient material between the
stellar wind termination shock and a second shock from the supersonic
motion.  The orientation of the bow shock arc depends on the direction
of the relative motion, although it can be skewed by density gradients
in the environment \citep{wilkin2000}.  The arc's standoff distance
from the star depends on the pressure balance between the stellar wind
and the ISM and hence on the magnitude of their relative motion and
the density of the ISM.  Bow-shock-like structures may also be
produced where dust in a photoevaporative flow is stalled by radiation
pressure rather than the stellar wind
\citep{ochsendorf2014a,ochsendorf2014b,ochsendorftielens2015,ochsendorf2015}.
The asymmetric stellar wind bubbles of slower-moving stars may also
have a similar appearance in the mid-infrared
\citep{mackey2015,mackey2016}.

Arc-shaped structures around massive stars have commonly been
considered a marker of high stellar velocities, under the assumption
that the relative motion between star and ISM is dominated by the
absolute motion of the star
\citep{vanburen1995,kaper1997,gvaramadzebomans2008,gvaramadze2010,kobulnicky2010,gvaramadze2011a,gvaramadze2011b}.
The typical velocity of an O-type star relative to its surroundings is
$\sim10$ km s$^{-1}$
\citep{blaauw1961,cruzgonzalez1974,giesbolton1986,tetzlaff2011},
comparable to the speed of sound in an \ion{H}{ii} region, but
20--30\% of O-type stars are ``runaways'' with velocities $\gtrsim40$
km s$^{-1}$
\citep{blaauw1961,cruzgonzalez1974,stone1991,tetzlaff2011}.  The high
speeds of runaway stars are imparted through dynamical interactions in a
cluster \citep{poveda1967,giesbolton1986,fujiiportegieszwart2011},
\btxt{through the disruption of a binary system when the companion star
  explodes as a supernova} \citep{blaauw1961}, or through a two-step
scenario involving both processes \citep{pflammaltenburgkroupa2010}.
Runaways make up 50--100\% of field O-type stars, the O-type
population found outside clusters and associations
\citep{dewit2005,schilbachroser2008,gvaramadze2012a}.

The question of whether all field O-type stars are runaways, or
whether a small fraction formed in isolation, is of key importance to
our understanding of massive star formation.  The monolithic collapse
model \citep{mckeetan2003,krumholz2005b,krumholz2009} permits truly
isolated massive star formation, albeit rarely, while the competitive
accretion model
\citep{zinnecker1982,bonnell2001a,bonnell2001b,bonnell2004} requires
that massive stars form exclusively in clusters.  In observational
studies, the presence of a bow shock \btxt{or candidate bow shock} is
sometimes taken as a clue that a given massive field star did not form
in situ.  For example, HD 48229 and HD 165319 were part of the
$4\pm2$\% of all O-type stars identified by
\citet{dewit2004,dewit2005} as likely candidates for isolated massive
star formation.  Bow shocks were later discovered around both sources
\citep{gvaramadzebomans2008,gvaramadze2012a}, calling into question
their origins in the field.

While 70\% of bow shocks \btxt{and bow-shock-like structures} are
located in relatively isolated environments consistent with runaway
stars \citep{kobulnicky2016}, the rest are found around OB stars in
clusters and associations.  These stars have sometimes been
interpreted as runaway interlopers from other regions
\citep{gvaramadze2011a}.  However, the assumption that the motion of
the ambient ISM is negligible relative to that of the star may not
always be valid, particularly in and around giant \ion{H}{ii} regions.
In many cases, bow shock orientations suggest that feedback-driven ISM
flows are relevant.  \citet{povich2008} observed that bow shocks in
the massive star-forming regions M17 and RCW 49 are oriented toward
those regions' central clusters, suggesting that global expansion of
the \ion{H}{ii} regions is the dominant component of the relative
star--ISM velocity.  Similarly, several bow shocks in Cygnus OB2 point
toward the association's interior \citep{kobulnicky2010}, as do more
than half of the candidate bow shocks in the Carina Nebula
\citep{smith2010b,sexton2015}.  The Galactic Plane survey of
\citet{kobulnicky2016} found that roughly 15\% of infrared bow shocks
are pointed at \ion{H}{ii} regions, while another 8\% face
bright-rimmed clouds; they also noted that bow shock orientations are
correlated on small scales, indicative of the influence of external
forces.  \citet{povich2008} refer to such feedback-facing bow shocks
as ``interstellar weather vanes,'' tracing \btxt{photoevaporative
  flows off local dense gas and/or large-scale gas motions driven by
  cluster feedback.}  \citet{kobulnicky2016} call them ``in-situ bow
shocks,'' reflecting their origin around presumably non-runaway OB
stars.

\btxt{When the motion of the star dominates over the motion of the
  surrounding ISM, as it does for runaway stars, the bow shock is
  expected to point in the direction of the star's motion.
  \citet{vanburen1995} surveyed bow shocks around known runaway stars
  and found that the bow shocks were preferentially aligned with their
  host stars' proper-motion vectors. However, they used proper motions
  measured in an absolute reference frame, not corrected for Galactic
  rotation and solar peculiar motion and thus not necessarily
  representative of a star's motion relative to the surrounding ISM.
  More recent surveys by \citet{peri2012,peri2015}, again of bow
  shocks around known runaways stars, did correct proper motions
  \ctxt{for Galactic rotation} and noted a similar, albeit
  qualitative, tendency for alignment.  Individual runaway stars are
  also often observed to be moving in the direction of their bow
  shocks \citep[e.g.,][]{moffat1998,moffat1999,comeronpasquali2007}.
  But what about bow shock around stars that have not already been
  identified as runaways?  \citet{kobulnicky2016} compiled a sample of
  bow shocks without any selection on their host stars' kinematics.
  They found that more than 50\% of the host stars with significant
  measured proper motions had velocity--bow shock misalignments of
  more than 45\degr, although again, they were working with absolute
  proper motions rather than local.  The relationship between stellar
  motion and bow shock orientation for stars in clusters and
  associations remains largely unexplored. }

\btxt{To further investigate this relationship,} we measure
\emph{local} proper motions for five \btxt{massive stars in the Carina
  Nebula (listed in Table \ref{tab:obstab}), each of which is
  associated with a candidate bow shock from \citet[][which includes
    objects first identified by \citealt{smith2010b}]{sexton2015}.
  \citet{smith2010b} and \citet{sexton2015} identified a total of 39
  ``extended red objects'' (EROs) in the Carina Nebula.}  These EROs
exhibit extended, often arc-shaped, morphology \btxt{in \emph{Spitzer}
  Infrared Array Camera (IRAC) 8.0 $\mu$m images.  Nine of the
  \citet{sexton2015} EROs are clearly resolved arcs and are classified
  as morphological bow shock candidates; one of our stars (ALS 15206)
  is associated with one of these sources (ERO 2).  Another eight of
  the \citet{sexton2015} EROs lack resolved morphologies at 8.0 $\mu$m
  but have infrared colors that rule out emission from young stellar
  objects (YSOs) and polycyclic aromatic hydrocarbons (PAHs).  The
  remaining four of our stars are associated with sources in this
  category, known as color bow shock candidates.}

%---------------------------------------------------------------------
\begin{table*}
  \caption{\emph{HST} data log.}
  \label{tab:obstab}
  \begin{tabular}{rlrrllllrlr}
    \hline
    ERO             & Star ID & R.A.    & Dec.    & Spectral & \ctxt{V}  & 
ACS       &
Date 1 & Exp. time 1 & Date 2 & Exp. time 2 \\
    No.$^{\mathrm{a}}$ &         & (J2000) & (J2000) & type     & \ctxt{(mag)} & 
    field &
       & (s)         &        & (s) \\
    \hline 
     2 & ALS 15206 & 10:44:00.9 & -59:35:46 & O9.2 V$^{\mathrm{b}}$ & 
    \ctxt{10.7$^{\mathrm{d}}$} & TR14 &
    2005 Jul 17 & 2 $\times$ 500  & 2015 Jun 28 & 2 $\times$
    520 \\ 
    23 & TYC 8626-2506-1 & 10:44:30.2 & -59:26:13 & O9 V$^{\mathrm{b}}$ & 
    \ctxt{10.9$^{\mathrm{e}}$} & TR14 &
    2005 Jul 17 & 2 $\times$ 500  & 2015 Jun 28 & 2 $\times$
    520 \\ 
    24 & CPD-59 2605 & 10:44:50.4 & -59:55:45 & B1 V$^{\mathrm{c}}$ & 
    \ctxt{11.1$^{\mathrm{f}}$} & POS27 &
    2006 Mar 18 & 2 $\times$ 500  & 2015 Mar 12 & 2 $\times$
    560 \\ 
    25 & HDE 305533 & 10:45:13.4 & -59:57:54 & B1 V$^{\mathrm{a}}$ & 
    \ctxt{10.6$^{\mathrm{f}}$} & POS26 & 
    2006 Mar 16 & 2 $\times$ 500  & 2015 Mar 12 & 2 $\times$ 
    560 \\ 
    31 & HD 93576 & 10:46:53.8 & -60:04:42 & O9.5 IV$^{\mathrm{b}}$ & 
    \ctxt{\phantom{1}9.6$^{\mathrm{d}}$} & POS20 & 
    2006 Mar 15 & 2 $\times$ 500  & 2015 Mar 11 & 2 $\times$ 
    455 \\ 
    \hline
  \end{tabular}
  \\
  \footnotesize
  \begin{tabular}{ll}
    $^{\mathrm{a}}$From \citet{sexton2015}. & 
    \ctxt{$^{\mathrm{d}}$From \citet{reed2003}.} \\
    $^{\mathrm{b}}$From \citet{sota2014}. & 
    \ctxt{$^{\mathrm{e}}$From \citet{hog2000}.} \\
    $^{\mathrm{c}}$From \citet{vijapurkardrilling1993}. & 
    \ctxt{$^{\mathrm{f}}$From \citet{masseyjohnson1993}.} \\
  \end{tabular}
\end{table*}
%---------------------------------------------------------------------

\btxt{Our five target stars reside in the Carina Nebula: their visual
  magnitudes \ctxt{(see Table \ref{tab:obstab})}, spectral types, and
  extinctions \citep{povich2011b} confirm that they are unlikely to be
  foreground or background objects.}  The Carina Nebula is home to
nearly 70 O-type and evolved massive stars \citep{smith2006a},
including some of the earliest known O-type stars \citep{walborn2002a}
and the luminous blue variable $\eta$ Carinae
\citep{davidsonhumphreys1997}.  At 2.3 kpc \citep{smith2006b}, it is
one of the closest and least-extincted massive star-forming regions.
Its two central clusters, Trumpler (Tr) 14 and Tr 16, contain about
half of its massive-star population.  The rest is spread across
$\sim$30 pc, mostly in a region of ongoing star formation known as the
South Pillars \citep{smith2000}. Emission-line profiles show that
feedback from the central clusters is driving the expansion of
multiple shells of ionized gas \citep{damiani2016}, resulting in a
global expansion of the \ion{H}{ii} region at $\pm$15--20 km s$^{-1}$
\citep{walbornhesser1975,walborn2002b,walborn2007}.  It is easy to
envision that the inward-facing orientations of many of Carina's
\btxt{candidate} bow shocks are the result of this supersonic,
feedback-driven ISM expansion \citep{sexton2015} or that they are
shaped by interaction with dense photoevaporative flows.  Here, we
explore whether those interpretations are valid and to what degree
these bow shocks are shaped by the motion and structure of the ISM
versus the motion of their driving stars.

The organization of this paper is as follows: In Section
\ref{sec:obs}, we describe our multiepoch \emph{Hubble Space
  Telescope} (\emph{HST}) observations, our image alignment procedure,
and our method for measuring proper motions.  We present our results
and compare the stellar motions to the orientations of their
associated \btxt{bow shock candidates} in Section
\ref{sec:results}. Section \ref{sec:disc} discusses the implications
and limitations of our results, and Section \ref{sec:conc} summarizes
our conclusions.

%%%%%%%%%%%%%%%%%%%%%%%%%%%%%%%%%%%%%%%%%%%%%%%%%%%%%%%%%%%%%%%%%%%%%%%
\section{Observations and Analysis}
\label{sec:obs}

\subsection{\emph{HST} ACS Imaging}
\label{subsec:hst}

\begin{figure*}
  \includegraphics[width=0.95\linewidth, trim=0 13mm 0 5mm,
    clip]{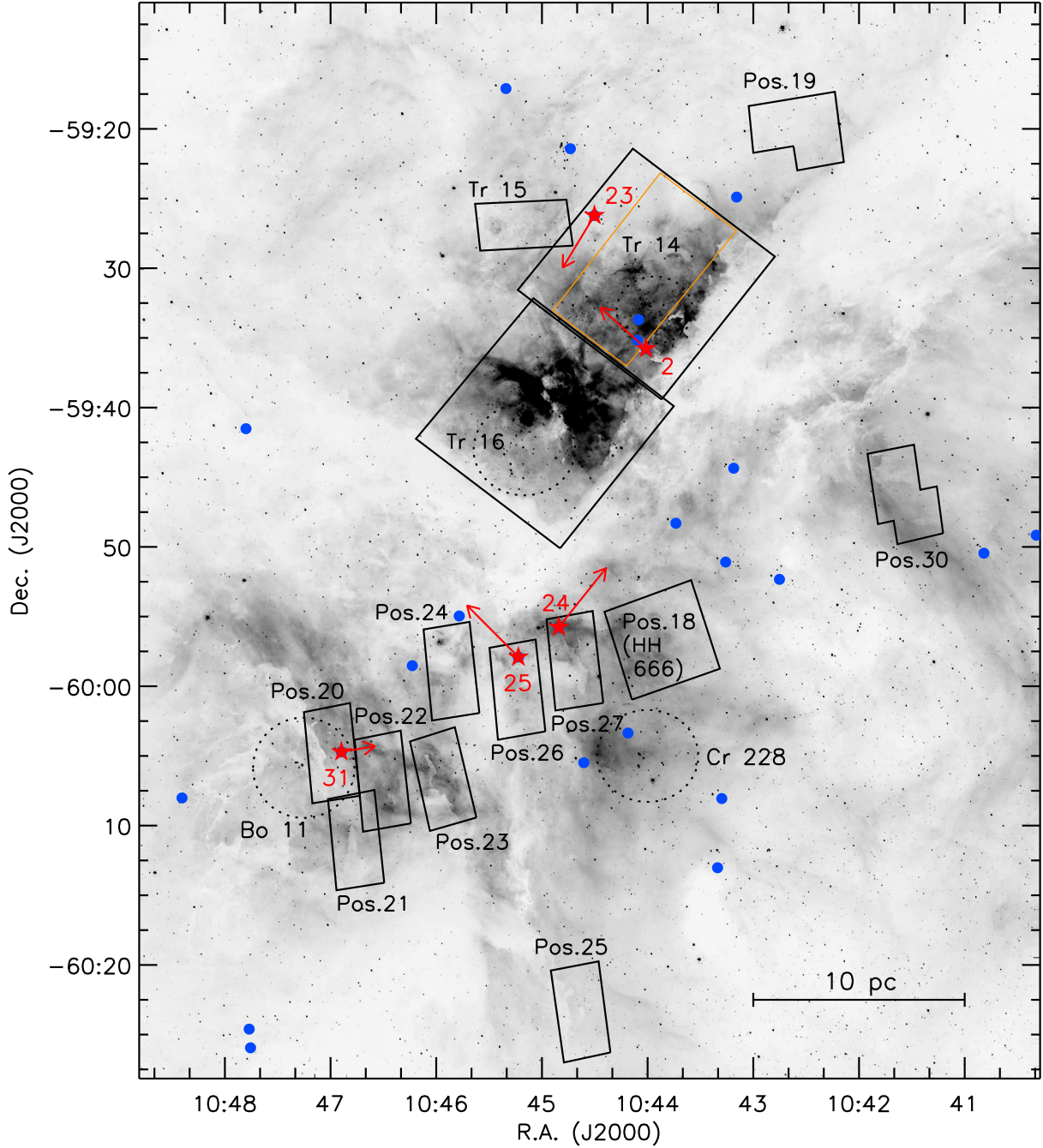}
  \caption{Ground-based H$\alpha$ image of the Carina Nebula from
    \citet{smith2010a}, with the positions of our \emph{HST} ACS
    fields outlined in black \ctxt{and labelled by their designations
      in the \emph{HST} data archive}.  All fields were observed
    twice, with a 9--10 year baseline between epochs.  The orange box
    indicates the portion of our Tr 14 mosaic that could not be
    re-observed at the same position angle due to changes in the
    \emph{HST} Guide Star Catalog between epochs.  Extended red
    objects (EROs) from \citet{sexton2015} that fall inside our fields
    are marked with red stars and labeled with their ERO number; the
    red arrows show their measured proper motions (Section
    \ref{sec:results}) scaled to a travel time of $10^5$ yr.  The blue
    filled circles are EROs from \citet{sexton2015} that fall outside
    our \emph{HST} fields.  The approximate positions of the primary
    central clusters, Tr 14 and Tr 16, are outlined with dashed
    circles, as are the positions of Bochum 11 and Collinder 228, two
    smaller clusters in the South Pillars region.}
  \label{fig:overview}
\end{figure*}

We have conducted a large-scale multiepoch survey of the Carina Nebula
using the Wide Field Camera (WFC) of \emph{HST}'s Advanced Camera for
Surveys (ACS).  All observations were made with the F658N filter,
which captures emission from H$\alpha$ and [\ion{N}{ii}] $\lambda$6584.
Our imaging coverage is shown in Figure \ref{fig:overview}, where each
small rectangle is one orbit made up of three overlapping pairs of
{\tt CR-SPLIT} exposures.  Orbital pointings were designed to target
features of particular interest in star formation (pillars,
\ctxt{Herbig-Haro objects}, etc.) as well as the central Tr 14 and Tr
16 clusters.  \ctxt{The pointings in Figure \ref{fig:overview} are
  labelled according to their designations in the \emph{HST} data archive.}

%---------------------------------------------------------------------
%---------------------------------------------------------------------

The first epoch of our ACS observations was taken in 2005--2006
(GO-10241 and GO-10475, PI: N. Smith; see \citealt{smith2010a}).  The
same set of observations was repeated in 2014--2015 (GO-13390 and
GO-13791, PI: N. Smith).  This second epoch was designed to replicate
the first as closely as possible in pointing and position angle in
order to minimize position-dependent systemic errors when measuring
proper motions.  Owing to changes in the \emph{HST} Guide Star Catalog
between epochs, we were unable to duplicate the orientation angle of
the central segment of the Tr 14 mosaic (marked in orange in Figure
\ref{fig:overview}) and of Positions 25 and 30.  Those observations
were rotated by $\sim$180\degr.

Also marked in Figure \ref{fig:overview} are the locations of EROs
from \citet{sexton2015}.  Our survey serendipitously imaged the stars
associated with seven EROs.  However, as discussed below, we were
unable to constrain the proper motions of the two ERO-associated stars in
the central part of the Tr 14 mosaic (the part for which the
orientation angle changed beween epochs), leaving us with a sample of
five.  Full details of the observations of each of these five stars
are given in Table \ref{tab:obstab}.

\subsection{Image Alignment and Stacking}
\label{subsec:align}
Our image alignment procedure, which adapts the methods of
\citet{anderson2008a,anderson2008b}, \citet{andersonvandermarel2010},
and \citet{sohn2012}, is described in detail in
\citet{reiter2015a,reiter2015b} and \citet{kiminki2016}.  In summary,
we find the positions of uncrowded, unsaturated stars in individual
exposures and use those positions to relate each image to a master,
distortion-free reference frame.  We use the program {\tt
  img2xym\_WFC.09x10} \citep{andersonking2006}, which uses an array of
effective point spread functions (PSFs) and has the option to fit a
spatially constant perturbation PSF to account for telescope breathing
and other focus changes.  The measured stellar positions were then
corrected for geometric distortion \citep{anderson2006}.

A master reference frame with a pixel scale of 50 mas was constructed
for each orbital pointing, aligned with north in the $+y$ direction.
The six overlapping images from each epoch of a given pointing were
stacked into two reference-frame master images (one per epoch) using
the stacking algorithm of \citet{anderson2008a}.  Object positions in
the master images are directly comparable between epochs to an
alignment accuracy of approximately 1 mas ($\sim$ 1 km s$^{-1}$ over a
9--10 year baseline at the distance of the Carina Nebula).  We found
that including a perturbation PSF in fitting stellar positions did not
improve the alignment precision, but we address other possible effects
of \emph{HST} focus changes in Section \ref{subsec:pms}.

In all cases, the master reference frames are not tied to an absolute
proper-motion zero point. Instead, the zero point is based on the
average motion of several hundred well-measured stars in the image.
In other words, the bulk motion of the Carina Nebula is removed, as
are smaller differences in the large-scale motion of Carina's clusters
and subclusters.  Features that are locally stationary, like bow
shocks, are expected to be stationary in our reference frames,
allowing direct measurement of the motion of stars relative to their
surroundings.

\subsection{Measuring Local Proper Motions of Saturated Stars}
\label{subsec:pms}

With the images from two epochs on the same reference frame, measuring
local proper motions for unsaturated stars in our stacked images is as
simple as comparing their PSF-derived positions between epochs.
However, most of the OB stars observed, including those associated
with \btxt{candidate} bow shocks, are saturated in
our ACS images (which were all $\sim$500 s long).  We were unable to
reconstruct the PSF core to perform traditional astrometry.  Instead,
we used the positions of the extended Airy rings, which are clearly
visible for these stars in these deep, high-resolution images.  The
left column of Figure \ref{fig:resid} shows the first-epoch image of
all five \btxt{ERO-associated} stars for which we measured proper
motions.

%---------------------------------------------------------------------
\begin{figure*}
  \includegraphics{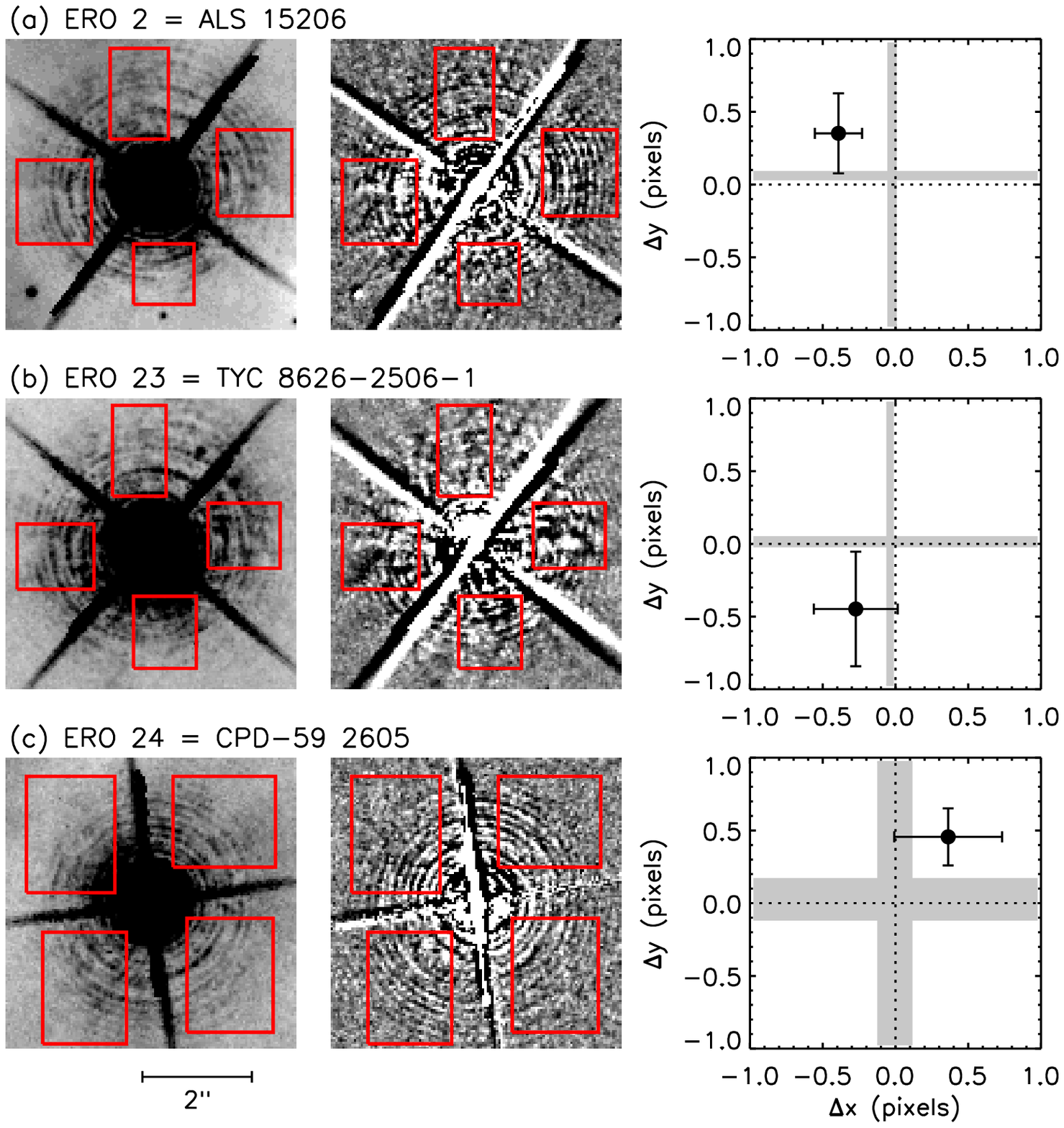}
  \caption{Left: first-epoch ACS F658N images of five stars associated
    with \btxt{candidate} bow shocks in the Carina Nebula.  The red
    boxes mark the sections of the Airy rings used for fitting the
    offset between epochs.  Middle: best-fit difference images
    (unshifted second epoch minus best-fit shifted first epoch) for
    each star.  Right: best-fit pixel offset between epochs with a
    9--10 year baseline.  The gray shaded regions mark the space of
    possible apparent offsets due to focus changes.}
  \label{fig:resid}
\end{figure*}
%---------------------------------------------------------------------
%---------------------------------------------------------------------
\begin{figure*}
  \includegraphics{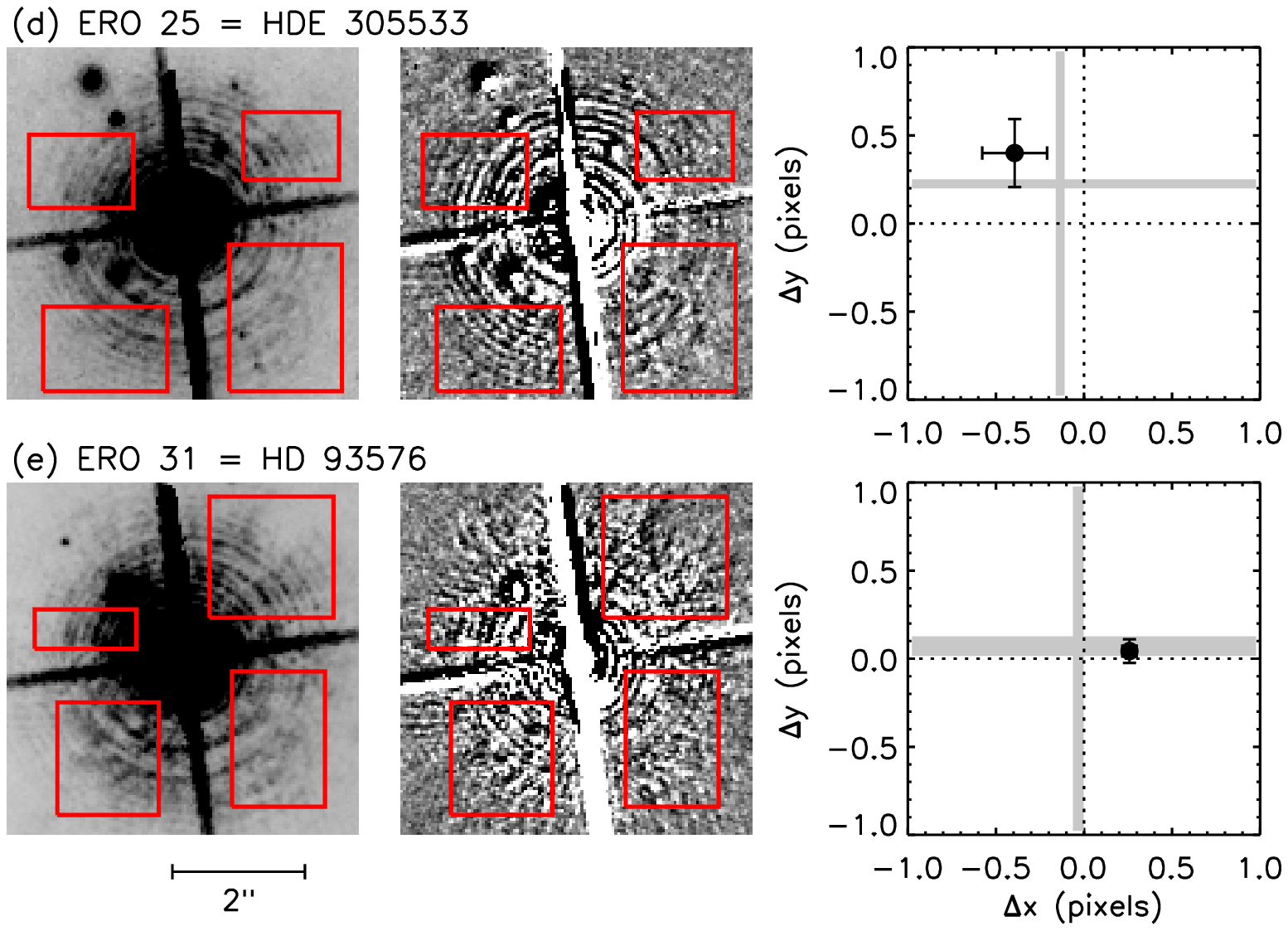} \contcaption{}
\end{figure*}
%---------------------------------------------------------------------

As marked in Figure \ref{fig:resid}, we identified four regions in the
outer PSF of each star, avoiding diffraction spikes, saturation
bleeding, and close companions.  We then found the pixel offset in $x$
(west-east) and $y$ (north-south) by which the first epoch needed to
be shifted in order to minimize the sum over those four regions of the
absolute value of the flux difference between epochs.  The best-fit
offset was computed using the {\tt AMOEBA} algorithm, the IDL
implementation of the downhill simplex function minimization method
\citep{neldermead1965,press1992}.  {\tt AMOEBA} requires an input
estimate.  As recommended by \citet{press1992}, we run the algorithm
twice, giving it a random starting estimate on the first run and then
starting the second run at the best-fit parameters of the first.  The
resulting best-fit difference images (unshifted second epoch minus
best-fit shifted first epoch) are shown in the middle column of Figure
\ref{fig:resid}.  Although the Airy rings do not disappear completely
in the difference images, there are no systematic differences in
residual flux between quadrants.  The best-fit offset for each star,
in pixels over the 9--10 year baseline, is plotted in the right column
of Figure \ref{fig:resid}.

This method does not produce accurate results for saturated stars in
fields that experienced significant rotation between epochs.  The
asymmetry of the ACS WFC PSF
\citep{andersonking2006,mahmudanderson2008} causes the flux
distribution in the outer PSF to be orientation-dependent.  When the
fields are rotated into alignment, the asymmetric flux distribution
introduces an apparent shift of up to several pixels.  Consequently,
as mentioned in Section \ref{subsec:hst}, we were unable to measure
the true shifts of the two ERO-associated stars in the central part of
our Tr 14 mosaic (see Figure \ref{fig:overview}).  These stars were
removed from further analysis and are not shown in Figure
\ref{fig:resid}; our final sample consists of the five ERO-associated
stars listed in Table \ref{tab:obstab}.

To characterize the uncertainties in our fits, we ran several
different tests.  First, we adjusted the size and placement of the
boxes used to calculate the residuals, then refit.  These adjustments
proved to have a negligible effect on the resulting best-fit offset.
Second, we repeated the full fitting process 100 times and measured
the standard deviation among the results.  These ranged from 0.03 to
0.4 pixels depending on the star.  Finally, we applied 100 random
[$x$,$y$] offsets to the first-epoch images and repeated the full
fitting process again for each, to determine how well we could recover
the expected (artificial $+$ true) offsets.  The standard deviation in
the offsets recovered via this approach ranged from 0.04 to 0.23
pixels.  For each star, we adopt the greater of the two uncertainty
values as the formal uncertainty.

Changes in \emph{HST} focus from both short-term thermal breathing and
long-term non-thermal effects \citep[e.g.,][]{coxlallo2012} could
induce an apparent offset between epochs by redistributing flux in the
PSF.  To evaluate the magnitude of this effect, we downloaded a Tiny
Tim model PSF \citep{krist2011} for each star, at its observed chip
position, for the appropriate focus value taken from the \emph{HST}
focus model \citep{dinino2008,niemilallo2010,coxniemi2011}.  The true
shift between ``epochs'' of Tiny Tim models is zero, so any measured
shift would be a false positive.  We ran the pair of Tiny Tim models
for each star through our fitting procedure and measured apparent
offsets of 0.035 to 0.22 pixels, which are illustrated by the shaded
gray regions in the third column of Figure \ref{fig:resid}.  In most
cases, the focus-induced shifts are small and/or distributed roughly
evenly about the origin.  For HDE 305533 (ERO 24), however, the focus
changes induced a systematic $-x, +y$ offset.  Removing this shift
would reduce the magnitude of the observed proper motion of HDE 305533
by roughly half, but would have only a small effect on its direction
of motion.

%%%%%%%%%%%%%%%%%%%%%%%%%%%%%%%%%%%%%%%%%%%%%%%%%%%%%%%%%%%%%%%%%%%%%%%
\section{Results}
\label{sec:results}

As the plots in the right column of Figure \ref{fig:resid}
demonstrate, each of the five \btxt{stars associated with candidate
  bow shocks} traveled no more than $\sim$0.5 pixels (25 mas) in any
direction over their 9--10 year baselines.  The measured pixel offsets
are given in Table \ref{tab:pmtab} along with the corresponding proper
motion components, the total transverse velocity, and the position
angle of the proper motion vector.  The best-fit local transverse
velocities range from 16 to 35 km s$^{-1}$; the red arrows in Figure
\ref{fig:overview} show the expected travel distances over 10$^5$ yr.
However, the uncertainties on most of the measured velocities are
relatively large: most of the stars have motion consistent with zero
within 1--2 $\sigma$.  Only HD 93576 has motion significant at the
3$\sigma$ level, in the $x$ direction, although it has negligible $y$
(north--south) motion.  We argue in Section \ref{subsec:up} below that
the true proper moions are likely on the smaller side of the allowed
ranges.  Even so, the results for all five stars constrain their
directions of motion to sectors less than 90\degr~wide.

%---------------------------------------------------------------------
\begin{table*}
  \caption{Local proper motions of stars associated with
    \btxt{candidate} bow shocks.}
  \label{tab:pmtab}
  \begin{tabular}{rlrrrrrr}
    \hline
    ERO No. & Star ID & $\delta x$ & $\delta y$ & $\mu_{\alpha}\cos\delta $  &
$\mu_{\delta}$   & v$_{\textrm{T}}$$^{\mathrm{a}}$ & Position angle  \\
            &         & (pixels)   & (pixels)   & (mas yr$^{-1}$)           &
(mas yr$^{-1}$) & (km s$^{-1}$)              & (deg E of N) \\
    \hline
 2 & ALS 15206       & -0.39 (0.16)  &  0.35 (0.28)  &  2.0 (0.8)  &  1.8 (1.4)  & 29 (17)  &  48 (25)  \\
23 & TYC 8626-2506-1 & -0.27 (0.29)  & -0.45 (0.39)  &  1.4 (1.4)  & -2.3 (2.0)  & 29 (27)  & 149 (35)  \\
24 & CPD-59 2605     &  0.36 (0.37)  &  0.46 (0.20)  & -2.0 (2.1)  &  2.5 (1.1)  & 35 (25)  & 322 (31)  \\
25 & HDE 305533      & -0.39 (0.18)  &  0.40 (0.19)  &  2.2 (1.0)  &  2.2 (1.1)  & 34 (16)  &  45 (19)  \\
31 & HD 93576        &  0.26 (0.04)  &  0.04 (0.07)  & -1.4 (0.2)  &  0.2 (0.4)  & 16 ( 5)  & 279 (15)  \\
    \hline
  \end{tabular}
  \\
  \footnotesize
  \begin{tabular}{l}
    Uncertainties for each quantity are listed in parentheses. \\
    $^{\mathrm{a}}$Total transverse velocity, assuming a distance of 2.3 kpc. \\ 
  \end{tabular}
\end{table*}

%---------------------------------------------------------------------

\begin{figure*}
  \includegraphics[width=0.85\linewidth]{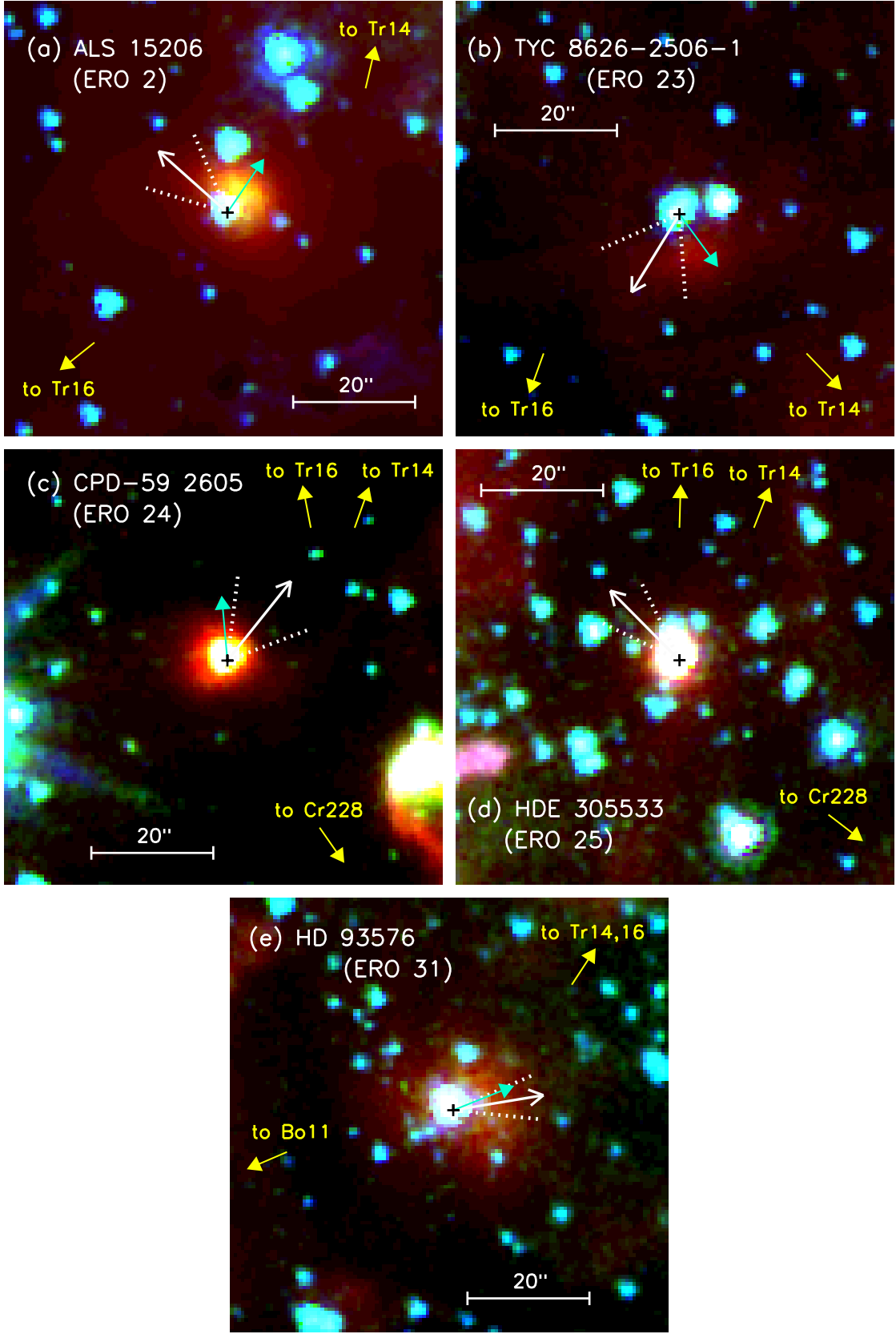}
  \caption{Composite three-color \emph{Spitzer} IRAC images of the
    five OB stars and their associated \btxt{candidate} bow shocks
    (blue = 3.6 $\micron$, green = 4.5 $\micron$, red = 8.0
    $\micron$).  The white arrows indicate the direction of the
    best-fit proper motion of each star, with the dotted white lines
    bracketing the $\pm1\sigma$ range of directions.  The cyan arrows
    highlight the orientation of the \btxt{candidate} bow shock, i.e.,
    the direction from the star to the peak of the 8.0 $\micron$
    emission, where measured by \citet{sexton2015}.  Also indicated
    are the directions to Tr 14 and Tr 16, the largest OB clusters in
    the Carina Nebula, and to the smaller clusters Bo 11 and Cr 228.}
  \label{fig:spitzer}
\end{figure*}

In Figure \ref{fig:spitzer}, we compare the local proper motions of
the stars to the orientations of their associated \btxt{candidate} bow
shocks.  The latter were determined by \citet{sexton2015} based on the
peaks of the 8.0 $\micron$ flux.  (ERO 25, associated with HDE 305533,
does not have a measured orientation.)  In these three-color
\emph{Spitzer} IRAC images, the \btxt{candidate} stellar wind bow
shocks appear as extended red (8.0 $\micron$) features, while nearby
stars are prominent in blue (3.6 $\micron$) and green (4.5 $\micron$).
We indicate the stars' motions with white arrows (lengths arbitrarily
scaled for visiblity) and show the range of possible directions with
dotted white lines.  The \btxt{orientations of the candidate bow
  shocks}, where known, are denoted by cyan arrows, and the outer
yellow arrows show the directions to the various OB clusters.

%---------------------------------------------------------------------
%---------------------------------------------------------------------

The uppermost panels in Figure \ref{fig:spitzer} show ALS 15206 (with
ERO 2 from \citealt{sexton2015}) and TYC 8626-2506-1 (ERO 23).  Both
of these stars are closer to Tr 14 than to Tr 16 (see Figure
\ref{fig:overview})\btxt{,} both \btxt{are associated with candidate}
bow shocks pointing at Tr 14, \btxt{and both have proper motions
  directed} tangentially to the orientation of their \btxt{candidate}
bow shocks.  \btxt{The radial velocity of ALS 15206 is poorly
  constrained, as it is a probable spectroscopic binary (Kiminki et
  al. in preparation), but is consistent with being drawn from the
  radial velocity distribution of Tr 14
  (\citealt{penny1993,garcia1998}; Kiminki et al. in preparation).  No
  radial velocity data exist for TYC 8626-2506-1.}  \btxt{Thus based
  on the proper motions of their associated stars, the} relative
motion shaping \btxt{EROs 2 and 23} appears to be dominated by the
motion of the surrounding ISM, expanding outward from Tr 14.  Unseen
density gradients may also play a role, but the motion of the stars
themselves do not look to be influencing the directions of \btxt{these
  candidate bow shocks}.  \btxt{They} may truly be acting as ``weather
vanes,'' tracing the large-scale flows of the ISM.

The middle row of Figure \ref{fig:spitzer} shows CPD-59 2605 (ERO 24)
and HDE 305533 (ERO 25).  These stars' \btxt{candidate} bow shocks are
not arc-shaped at IRAC resolutions; \citet{sexton2015} were able to
measure an orientation for ERO 24 but not for ERO 25.  Both stars are
in Carina's South Pillars region, and are roughly 7.5 pc northeast of
the nominal center \citep{wu2009} of the sparse open cluster Cr
228. ERO 24 points north toward Tr 16, suggesting that it is
influenced by feedback-driven outflows.  However, its associated star
(CPD-59 2605) has a local proper motion to the northwest, consistent
with the \btxt{orientation of the candidate bow shock} within the
uncertainties.  It is thus not possible to distinguish between the
effects of ISM flows and stellar motion in the case of ERO 24, as both
may be relevant to shaping that \btxt{feature}.  \btxt{No radial velocity
  data exist for CPD-59 2605.}

Relative to its surroundings, HDE 305533 (ERO 25) is moving to the
northeast, away from the WNH star HD 93131 and the small group of late
O-type stars that make up the center of Cr 228.  Its path hints at an
ejection from Cr 228, although at its observed speed it would have
covered the 7.5 pc from Cr 228 in just 220,000 yr (but see discussion
in Section \ref{subsec:up} below on the likelihood that our measured
proper motions are upper limits).  The age and extent of Cr 228 are
also poorly constrained, as it has often been considered an extension
of Tr 16 \citep{walborn1995,smithbrooks2008}, while X-ray data show it
to be a discrete collection of groups and subclusters without a clear
center \citep{feigelson2011}.  The origin of HDE 305533 is therefore
not clearly evident.  \btxt{Its radial velocity \citep[-18 km
    s$^{-1}$;][]{levato1990} is typical for the massive stars in Cr
  228 and the South Pillars region (\citealt{levato1990}; Kiminki et
  al. in preparation) and comparable to the radial velocity of the
  surrounding gas pillars \citep{rebolledo2016}.}

Finally, the bottom panel of Figure \ref{fig:spitzer} shows HD 93576,
the binary system \citep{levato1990} associated with ERO 31.  HD 93576
lies on the outskirts of the small open cluster Bochum 11 (Bo 11),
located in the southeastern part of the South Pillars.  Bo 11 is home
to an estimated 1000 stars \citep{dias2002}, including the O5
supergiant HD 93632 \citep{sota2014} and three additional O-type stars
besides HD 93576.  Photometric analysis indicates that the cluster is
3--5 Myr old
\citep{fitzgeraldmehta1987,patatcarraro2001,preibisch2011c}; the
presence of an O5I star suggests that 3 Myr is more likely.  As Figure
\ref{fig:spitzer} shows, the proper motion vector of HD 93576 is
closely aligned with the orientation of its \btxt{candidate} bow
shock, which in turn points nearly directly away from the center of Bo
11.  This configuration suggests that HD 93576 was ejected from Bo 11
and that its subsequent supersonic motion produced the observed
\btxt{candidate} bow shock.  \btxt{Its systemic radial velocity (-8 km
  s$^{-1}$; Kiminki et al. in preparation) is commensurate with the
  radial velocities of the other massive members of Bo 11
  (\citealt{levato1990}; Kiminki et al. in preparation) and the nearby
  dense gas \citep{rebolledo2016}.}  But its observed proper motion
(15 km s$^{-1}$) and current position (1.9 pc from the center of Bo
11) indicate an ejection date just 130,000 yr ago.  Perhaps HD 93576
was ejected 2--3 Myr after the formation of Bo 11 \citep[possible;
  see][]{ohkroupa2016}.  Alternately, it could have originated outside
the Carina Nebula \btxt{and have a coincidental agreement in radial
  velocities}, or the magnitude of its proper motion could be smaller
than measured (see discussion in Section \ref{subsec:up}).  In
addition, its \btxt{candidate} bow shock is also generally directed
toward the interior of the Carina Nebula, so a contribution from ISM
flows driven by cluster feedback cannot be ruled out regardless of the
origin of the star.

%%%%%%%%%%%%%%%%%%%%%%%%%%%%%%%%%%%%%%%%%%%%%%%%%%%%%%%%%%%%%%%%%%%%%%%
\section{Discussion}
\label{sec:disc}

\subsection{Proper Motions as Upper Limits}
\label{subsec:up}

For four of the five \btxt{candidate} bow shock host stars in our
sample, we measure local proper motions of $\sim$30 km s$^{-1}$, with
associated uncertainties of $>15$ km s$^{-1}$.  (This total includes
HDE 305533, whose observed motion may include a contribution from
focus changes as described in Section \ref{subsec:pms}.)  Several
lines of reasoning support the interpretation of these measured
velocities as upper limits, with the true proper motions lying on the
small side of the allowed range.  First, the typical velocity of an
O-type star relative to its surrounding is $\sim$ 10 km s$^{-1}$
\citep{blaauw1961,cruzgonzalez1974,giesbolton1986,tetzlaff2011}.  Of
course, the stars in our sample are arguably not typical, given their
association with \btxt{candidate} dusty bow shocks.  Space velocities
of 30 km s$^{-1}$ may qualify them as runaway stars, depending on the
choice of runaway classification criteria.  None of the five stars
measured here are moving with trajectories that could have originated
in Tr 14 or Tr 16, although HDE 305533 (ERO 25) and HD 93576 (ERO 31)
may have come from the smaller open clusters Cr 228 and Bo 11,
respectively.  An object moving at 30 km s$^{-1}$ would cover 60 pc in
2 Myr \citep[the average estimated age of Tr
  14/16;][]{walborn1995,smith2006a,preibisch2011c}, and these stars
are all significantly closer than that to any possible clusters of
origin in the Carina Nebula.  It is possible that all four of the
stars with measured proper motions of $\sim$30 km s$^{-1}$ were
ejected more recently, but that scenario would still not explain
their directions of motion.  Similarly, it is possible that all four
are interlopers in the Carina Nebula, originating from another
cluster, but the chance of encountering four such stars in our small
sample is low.  \btxt{And as described in Section \ref{sec:results},
  the radial velocities of our sample stars, where available, agree
  with the radial motions of the surrounding stars and gas, consistent
  with more local origins.}

In addition, speeds of 30 km s$^{-1}$ are inconsistent with the
relative star--ISM velocities computed for Carina's \btxt{candidate}
bow shocks by \citet{sexton2015}.  The pressure balance governing a
standard bow shock makes it possible to estimate the relative
star--ISM velocity as a function of measured standoff distance by
making reasonable assumptions about stellar wind velocity, mass-loss
rate, and ISM density.  \citet{sexton2015} measured the standoff
distances of nine EROs in the Carina Nebula and found an average
star--ISM velocity of 17 km s$^{-1}$.  For ERO 2 (associated with ALS
15206), the relative star--ISM velocity was a barely-supersonic 7 km
s$^{-1}$.  Similar relative velocities for bow shocks in the massive
star-forming region RCW 38 were reported by \citet{winston2012}.
These numbers have substantial uncertainties due to the assumptions
that go into their calculation, but they still suggest somewhat lower
stellar velocities.  Consider ERO 2 (ALS 15206): The orientation of
the \btxt{candidate} bow shock indicates that the direction of the
highest relative star--ISM velocity is to the northwest.  We have
measured that ALS 15206 is moving to the northeast, tangential to its
\btxt{candidate} bow shock.  If the relative star--ISM velocity in the
direction of the \btxt{candidate} bow shock is on the order of 7 km
s$^{-1}$, the relative star--ISM velocity in a different direction
cannot be substantially higher than that, although the picture may be
complicated if there are density gradients in the ISM.

For these reasons, it is unlikely that the measured \btxt{stars
  associated with candidate bow shocks} are moving as fast as 30 km
s$^{-1}$ relative to their surroundings.  The local proper motions
given here should thus be treated as upper limits.  HD 93576 may be an
exception, as its westward motion is measured at 3$\sigma$
significance (but this raises questions about its possible origin in
Bo 11, as discussed in Section \ref{sec:results}).

\subsection{Comparison to Absolute Proper Motions}
\label{subsec:abs}

\begin{figure*}
  \includegraphics{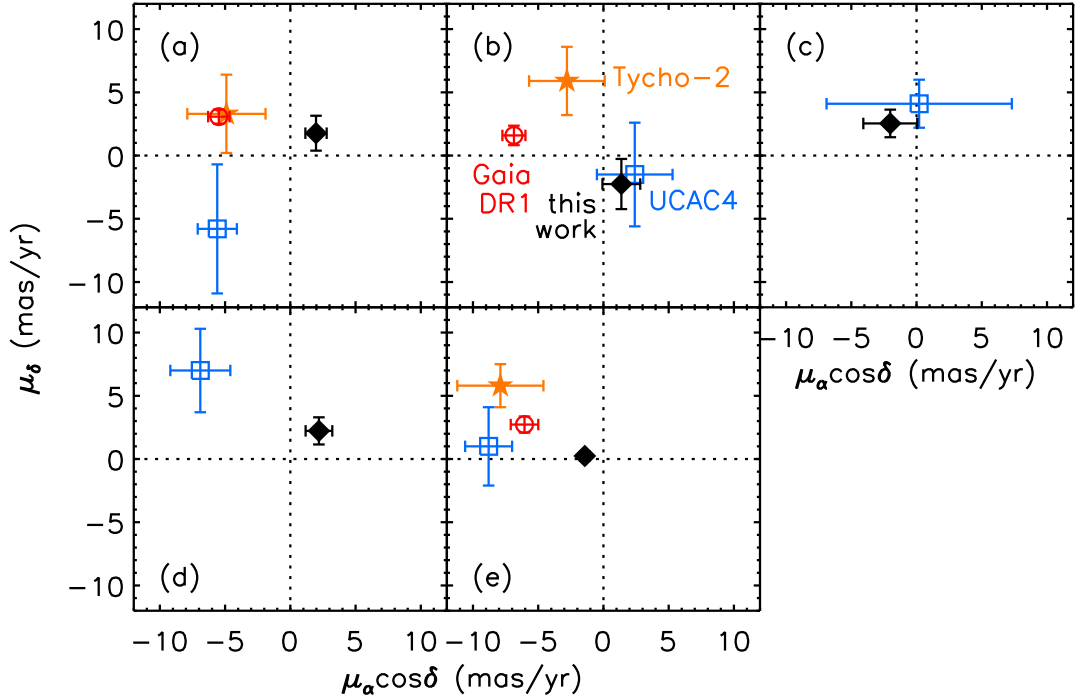}
  \caption{Comparison of the \emph{local} proper motion (filled
    black diamonds) of each star measured here to its \emph{absolute} proper
    motion from UCAC4 \citep[open blue squares,][]{zacharias2013},
    the Tycho-2 Catalogue \citep[filled orange stars,][]{hog2000}, and
    \emph{Gaia} Data Release 1 \citep[open red
      circles,][]{gaia2016a,gaiabrown2016,lindegren2016}.  (a) ALS
    15206 (ERO 2); (b) TYC 8626-2506-1 (ERO 23); (c) CPD-59 2605 (ERO
    24); (c) HDE 305533 (ERO 25); (e) HD 93576 (ERO 31).}
  \label{fig:comp}
\end{figure*}

\btxt{All five of the stars in our sample have proper motions listed
  in the USNO CCD Astrograph Catalog \citep[UCAC4;][]{zacharias2013},
  and ALS 15206, TYC 8626-2506-1, and HD 93576 also have proper
  motions in the Tycho-2 Catalogue \citep{hog2000} and \emph{Gaia}
  Data Release 1
  \citep[DR1;][]{gaia2016a,gaiabrown2016,lindegren2016}.  (Note that
  Tycho-2 and \emph{Gaia} DR 1 are not wholly independent
  measurements, as the latter incorporates positional information from
  the former.)  The UCAC4, Tycho-2, and \emph{Gaia} DR1 proper motions
  are measured in an absolute reference frame and are therefore not
  directly comparable to the local proper motions measured here.  We
  would expect to see a roughly constant offset between these absolute
  proper motions and our local ones, with that offset representing the
  bulk motion of the Carina Nebula relative to the Sun.  We plot the
  available absolute proper motions for each star, along with our
  measured local proper motions, in Figure \ref{fig:comp}.  Contrary
  to expectations, Figure \ref{fig:comp} does not show a
  consistent offset between local and absolute proper motions.  The
  UCAC4 proper motions in particular do not follow any apparent trend
  relative to the local proper motions or the Tycho-2 and \emph{Gaia}
  DR1 data.  The differences between catalogues suggest that there may
  be systematic effects in the literature measurements that are not
  taken into account in the published uncertainties.}

%---------------------------------------------------------------------
%---------------------------------------------------------------------

\begin{figure*}
  \includegraphics{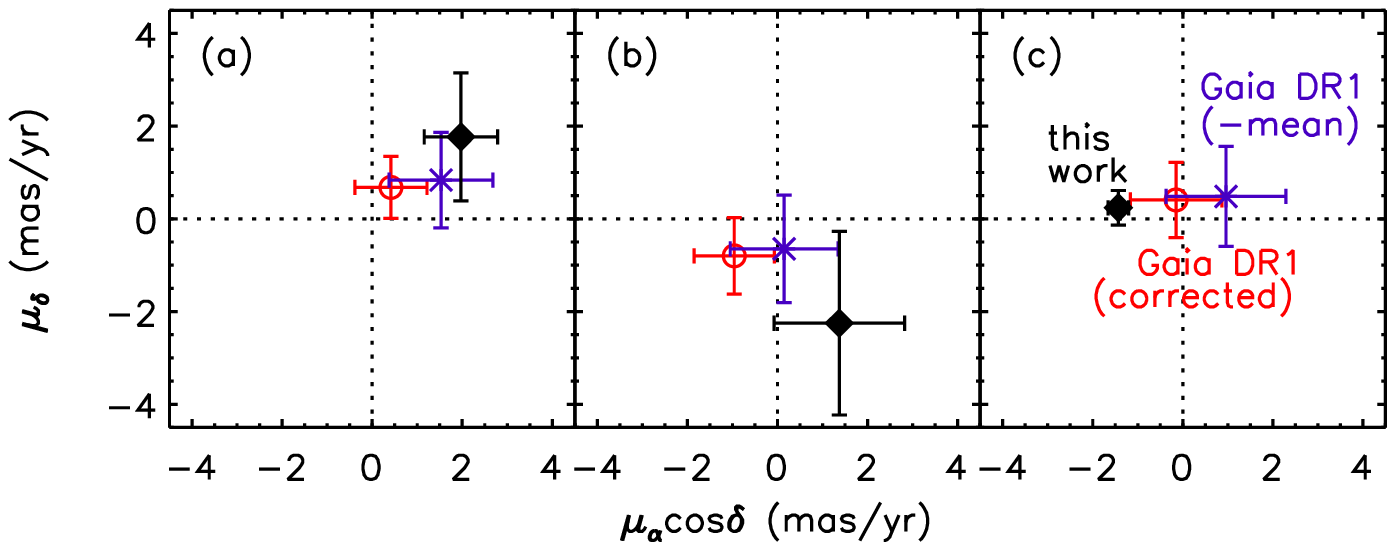}
  \caption{\ctxt{Comparison of the local proper motions (filled black
      diamonds) measured here to proper motions from \emph{Gaia} Data
      Release 1 \citep{gaia2016a,gaiabrown2016,lindegren2016} that
      have been corrected to a local reference frame.  Open red
      circles are the \emph{Gaia} DR1 data corrected for Galactic
      rotation and solar peculiar motion as described in Section
      \ref{subsec:abs}; purple asterisks are the \emph{Gaia} DR1 data
      corrected empirically by subtracting the mean proper motion of
      O-type stars in the Carina Nebula.  (a) ALS 15206 (ERO 2); (b)
      TYC 8626-2506-1 (ERO 23); (c) HD 93576 (ERO 31).}}
  \label{fig:gaia}
\end{figure*}

\ctxt{For further comparison, we correct the \emph{Gaia} proper
  motions, where available, to the rest frame of the Carina Nebula in
  two ways.  First, we formally correct for Galactic rotation and
  solar peculiar motion, as in \citet{moffat1998,moffat1999} and
  \citet{comeronpasquali2007}.  We adopt Oort's constants $A=15\pm1$
  km s$^{-1}$ kpc$^{-1}$ and $B=-12\pm1$ km s$^{-1}$ kpc$^{-1}$
  \citep{feastwhitelock1997,elias2006,bovy2017} and components of the
  solar peculiar velocity
  $(U_{\sun},V_{\sun},W_{\sun})=(10\pm1,12\pm1,7\pm1)$ km s$^{-1}$
  \citep{feastwhitelock1997,elias2006,schonrich2010,tetzlaff2011}.
  The corrected proper motions are plotted in Figure \ref{fig:gaia}.
  For all three stars, the corrected \emph{Gaia} proper motions are
  $\le1.3$ mas yr$^{-1}$ ($\le14$ km s$^{-1}$), supporting our
  interpretation that these stars are not runaways.  The corrected
  \emph{Gaia} proper motion of ALS 15206 (ERO 2) is, like our measured
  motion, directed to the northeast, tangential to the orientation of
  its candidate bow shock.  The corrected \emph{Gaia} motion of TYC
  8626-2506-1 (ERO 23) is to the southwest, into its candidate bow
  shock, although its 1$\sigma$ uncertainties overlap with those of
  our measured motion to the southeast. The corrected \emph{Gaia}
  motion of HD 93576 (ERO 31) is also consistent with our data,
  although the \emph{Gaia} results indicate a smaller velocity to the
  west, suggesting a longer time since ejection from Bo 11.

  We also perform an empirical correction to the local reference
  frame: we compute the weighted mean proper motion of the 38 O-type
  stars in the Carina Nebula in \emph{Gaia} DR1 \citep[which is
    roughly half the total O-type population of the region;
    e.g.,][]{smith2006a,gagne2011,alexander2016} and subtract that from
  the absolute \emph{Gaia} proper motions of the three sample stars.
  The results are consistently $\sim1.1$ mas yr$^{-1}$ ($\sim12$ km
  s$^{-1}$) offset from the results of formally correcting for
  Galactic rotation and solar peculiar motion.  For ALS 15206 (ERO 2)
  and TYC 8626-2560-1 (ERO 23), the empirical correction brings the
  corrected \emph{Gaia} proper motions into better agreement with our
  results.  For HD 93576 (ERO 31), the empirical correction produces
  worse agreement with our results and suggests that the star is
  moving to the east, away from its candidate bow shock and toward Bo
  11.  The different parts of the Carina Nebula may have different
  large-scale motions not properly accounted for in these corrections.
  Future \emph{Gaia} data releases, extending into Carina's
  intermediate-mass population, will allow more precise and
  locally-specific corrections to the local reference frame.  }

%---------------------------------------------------------------------
%---------------------------------------------------------------------

\subsection{Interpreting Bow Shocks in Giant \ion{H}{ii} Regions}
\label{subsec:interp}

In our subsample of bow shock candidates in the Carina Nebula, EROs 2
and 23 face the OB cluster Tr 14, while EROs 24 and 31 point more
generally toward Tr 14 and 16.  The majority of bow shock candidates
in the full \citet{sexton2015} sample also point in toward the
clusters rather than out as would be expected for runaway stars.
\citet{sexton2015} hypothesized that these \btxt{candidate} bow shocks
are markers of large-scale ISM flows driven by cluster feedback.  The
ionized gas in the Carina Nebula is known to be globally expanding at
15--20 km s$^{-1}$ \citep{walbornhesser1975,walborn2002b,walborn2007},
with multiple local centers of expansion including Tr 14
\citep{damiani2016}.  Feedback-driven outflows are also thought to
explain the orientations of bow shocks in other massive star-forming
regions \citep{povich2008,winston2012} and the correlation of bow
shock orientations on small angular scales \citep{kobulnicky2016}.
Our results are broadly compatible with this interpretation, but they
indicate that the factors influencing a bow shock \btxt{or
  bow-shock-like structure} cannot be deduced solely from its
orientation.

The associated stars of EROs 2 and 23 are not moving in the direction
of their \btxt{infrared} arcs, which are thus likely shaped by
feedback from Tr 14.  These two objects confirm that, at least in this
environment, \btxt{apparent} bow shock orientation does not always
follow stellar motion.  In contrast, the associated stars of EROs 24,
25, and 31 are moving roughly toward Tr 14 and 16.  Both stellar
motion and ISM flows could be relevant in setting the orientation of
these three \btxt{candidate} bow shocks, demonstrating that
cluster-facing \btxt{bow-shock-like structures} are not necessarily
clear markers of the motion of the ISM.

Feedback may also be affecting bow shocks in giant \ion{H}{ii} regions
in other ways besides large-scale outflows.  For instance, the pile-up
of dust in bow shock arcs depends on the presence of dust in the
surrounding \ion{H}{ii} region, as the hot winds of OB stars do not
make dust effectively.  This dust may originate in photoevaporative
flows off nearby molecular pillars, driven globally by ionizing
radiation from the central clusters and locally by individual OB
stars.  \citet{kobulnicky2016} found that eight percent of bow shocks
across the Galactic Plane face bright-rimmed clouds, suggesting they
are shaped by local photoevaporative flows.  The arc-shaped dust waves
around $\sigma$ and $\lambda$ Ori are also thought to be driven by
photoevaporative flows off the edge of ionized bubbles
\citep{ochsendorf2014a,ochsendorf2014b,ochsendorftielens2015,ochsendorf2015}.
Density gradients in the ISM can also affect bow shock symmetries
\citep{wilkin2000} or create infrared arcs via uneven heating.

We inspected \emph{Spitzer} images of the Carina Nebula to assess the
relationship between Carina's EROs, its molecular gas, and the
distribution of warm dust.  There is a possible tendency for EROs to
be closer to dense pillars than expected from a random distribution,
but there is no correlation between ERO orientation and the direction
to the nearest pillar.  ERO 31, for example, lies just
45\arcsec~($\sim$0.5 pc) from the edge of a prominent pillar, but
points almost directly away from it.  Multiband Imaging Photometer
(MIPS) images at 24 $\micron$
reveal complex warm dust structures throughout the nebula, including
around EROs 24 and 31.  However, the origin and impact of these
structures with respect to the EROs is unclear.  Higher-resolution
mid-infrared imaging is required to tease out the effects of density
gradients and photoevaporative flows in shaping Carina's EROs.

In any case, our main result is unaffected: in a giant \ion{H}{ii}
region, \btxt{the orientation of bow-shock-like structures} may be
determined by the ISM, by stellar motion, or by some combination of
factors.  It is worth reiterating that none of the five \btxt{stars}
in our study are runaways from Tr 14 or 16.  While this result is
unsurprising given the orientation of their \btxt{candidate} bow
shocks, it confirms that stars with \btxt{bow-shock-like structures}
are not automatically runaways.  The statistical preference for
alignment between stellar motion and bow shock orientation,
particularly among known runaway stars
\citep{vanburen1995,kobulnicky2016} suggests that stellar motion does
dominate over ISM flows for bow shocks far from feedback-generating
clusters.  But within associations, assumptions about the implications
of bow shocks \btxt{and bow-shock-like structures}
\citep[e.g.,][]{kobulnicky2010,gvaramadze2011a} should be made with
caution.

\subsection{Implications for the Origins of OB Associations}
\label{subsec:assoc}

The local proper motions of CPD-59 2605 (ERO 24), HD 305533 (ERO 25),
and HD 93576 (ERO 31) can also shed light on the origins of the
distributed massive-star population in the Carina Nebula.  Nearly half
of Carina's massive stars, including the WNH star HD 93131, are spread
across roughly 20 pc in the South Pillars \citep{smith2006a}.  Some of
these massive stars are associated with small open clusters (Bo 11, Cr
228) and other groups and subclusters of young stars
\citep{smith2010b,feigelson2011}.  However, \emph{Herschel} imaging
detected no massive protostars in the region, suggesting that the
ongoing star formation in the South Pillars is limited to low- and
intermediate-mass stars \citep{gaczkowski2013}.  

In the classic picture of clustered star formation
\citep[e.g.,][]{ladalada2003}, massive stars rarely form in a
distributed mode as seen in the South Pillars.  Instead, massive stars
are born in clusters that may subsequently become unbound after gas
dispersal and expand into OB associations
\btxt{\citep{tutukov1978,hills1980,ladalada1991,ladalada2003}}.  In this
picture, one would expect the Carina Nebula's distributed massive
stars to have formed in the central Trumpler clusters and drifted out
to their current locations over several Myr.

Our proper motion results are inconsistent with this expectation, as
all three of the massive South Pillars stars measured here are moving
toward the Trumpler clusters, not away.  These stars' kinematics
suggest that they were born in the South Pillars, possibly in one of
the smaller open clusters, and support a model of star formation in
which OB associations form directly as loose aggregates
\citep[e.g.,][]{efremovelmegreen1998b,clark2005}.  A similar result
has been observed for the Cyg OB2 association based on its
substructure and lack of global expansion
\citep{wright2014,wright2016}.  Further investigation of stellar
kinematics in the South Pillars is needed to confirm this
interpretation of the Carina Nebula's distributed population.

%%%%%%%%%%%%%%%%%%%%%%%%%%%%%%%%%%%%%%%%%%%%%%%%%%%%%%%%%%%%%%%%%%%%%%%
\section{Conclusions}
\label{sec:conc}

Using \emph{HST} ACS imaging with 9--10 year baselines, we have
measured the local proper motions (i.e., relative to the surrounding
stars) of five OB stars associated with candidate bow shocks in the
Carina Nebula.  Because these stars are highly saturated in our data,
we use precisely-aligned images to measure the shift in each star's
Airy rings between epochs.  The results are largely upper limits, but
we are able to constrain the direction of each star's motion for
comparison to the orientation of its \btxt{candidate} bow shock.

Stellar wind bow shocks are formed when the relative velocity between
star and ISM is supersonic, but the bow shock alone does not indicate
which component of the relative velocity dominates.  Are bow shocks
indicators of fast-moving runaway stars or do they mark the
large-scale flow of the ISM?  In our sample of five, we find two cases
where the latter is likely the case, as the stars are moving at a
tangent to the arc of their \btxt{candidate} bow shocks.  In the other
three cases, we conclude that the possible influences of ISM flows,
ISM structure, and stellar motion cannot be separated, and that
multiple factors could be relevant for each object.  We consequently
caution against overinterpreting the orientation of bow shocks
\btxt{and bow-shock-like structures} in giant \ion{H}{ii} regions like
the Carina Nebula.

In addition, none of the five stars measured here are runaways from
the central OB clusters of the Carina Nebula, although two may have
been ejected from smaller open clusters in the South Pillars.  This
finding emphasizes that bow shocks \btxt{and bow-shock-like
  structures} in giant \ion{H}{ii} regions are not definite markers of
runaway stars.  It also suggests that the distributed massive-star
population in the Carina's South Pillars formed along with the
distributed low- and intermediate-mass population; the resulting OB
association is not the expanding remnant of an embedded cluster but a
loose collection of many small groups and clusters.

%%%%%%%%%%%%%%%%%%%%%%%%%%%%%%%%%%%%%%%%%%%%%%%%%%%%%%%%%%%%%%%%%%%%%%%
\section*{Acknowledgements}

The authors would like to thank Jay Anderson for providing us with his
suite of PSF-fitting and image alignment software, and for his
valuable instruction, guidance, and technical support.  \btxt{We also
  thank the anonymous referee for a constructive review.}  Support for
this work was provided by NASA grants GO-13390 and GO-13791 from the
Space Telescope Science Institute, which is operated by the
Association of Universities for Research in Astronomy, Inc. under NASA
contract NAS 5-26555.  This work is based on observations made with
the NASA/ESA \emph{Hubble Space Telescope}, obtained from the Data
Archive at the Space Telescope Science Institute.  This work has made
use of data from the European Space Agency (ESA) mission {\it Gaia}
(\url{http://www.cosmos.esa.int/gaia}), processed by the {\it Gaia}
Data Processing and Analysis Consortium (DPAC,
\url{http://www.cosmos.esa.int/web/gaia/dpac/consortium}). Funding for
the DPAC has been provided by national institutions, in particular the
institutions participating in the {\it Gaia} Multilateral Agreement.

%%%%%%%%%%%%%%%%%%%%%%%%%%%%%%%%%%%%%%%%%%%%%%%%%%%%%%%%%%%%%%%%%%%%%%%
\bibliographystyle{mnras}
\bibliography{ms} 

\begin{thebibliography}{}
\makeatletter
\relax
\def\mn@urlcharsother{\let\do\@makeother \do\$\do\&\do\#\do\^\do\_\do\%\do\~}
\def\mn@doi{\begingroup\mn@urlcharsother \@ifnextchar [ {\mn@doi@}
  {\mn@doi@[]}}
\def\mn@doi@[#1]#2{\def\@tempa{#1}\ifx\@tempa\@empty \href
  {http://dx.doi.org/#2} {doi:#2}\else \href {http://dx.doi.org/#2} {#1}\fi
  \endgroup}
\def\mn@eprint#1#2{\mn@eprint@#1:#2::\@nil}
\def\mn@eprint@arXiv#1{\href {http://arxiv.org/abs/#1} {{\tt arXiv:#1}}}
\def\mn@eprint@dblp#1{\href {http://dblp.uni-trier.de/rec/bibtex/#1.xml}
  {dblp:#1}}
\def\mn@eprint@#1:#2:#3:#4\@nil{\def\@tempa {#1}\def\@tempb {#2}\def\@tempc
  {#3}\ifx \@tempc \@empty \let \@tempc \@tempb \let \@tempb \@tempa \fi \ifx
  \@tempb \@empty \def\@tempb {arXiv}\fi \@ifundefined
  {mn@eprint@\@tempb}{\@tempb:\@tempc}{\expandafter \expandafter \csname
  mn@eprint@\@tempb\endcsname \expandafter{\@tempc}}}

\bibitem[\protect\citeauthoryear{{Alexander}, {Hanes}, {Povich}  \&
  {McSwain}}{{Alexander} et~al.}{2016}]{alexander2016}
{Alexander} M.~J.,  {Hanes} R.~J.,  {Povich} M.~S.,   {McSwain} M.~V.,  2016,
  \mn@doi [\aj] {10.3847/0004-6256/152/6/190}, \href
  {http://adsabs.harvard.edu/abs/2016AJ....152..190A} {152, 190}

\bibitem[\protect\citeauthoryear{{Anderson}}{{Anderson}}{2006}]{anderson2006}
{Anderson} J.,  2006, in {Koekemoer} A.~M.,  {Goudfrooij} P.,   {Dressel}
  L.~L.,  eds, The 2005 HST Calibration Workshop: Hubble After the Transition
  to Two-Gyro Mode. p.~11

\bibitem[\protect\citeauthoryear{{Anderson} \& {King}}{{Anderson} \&
  {King}}{2006}]{andersonking2006}
{Anderson} J.,  {King} I.~R.,  2006, Technical report, {PSFs, Photometry, and
  Astronomy for the ACS/WFC}.
Space Telescope Science Institute

\bibitem[\protect\citeauthoryear{{Anderson} \& {van der Marel}}{{Anderson} \&
  {van der Marel}}{2010}]{andersonvandermarel2010}
{Anderson} J.,  {van der Marel} R.~P.,  2010, \mn@doi [\apj]
  {10.1088/0004-637X/710/2/1032}, \href
  {http://adsabs.harvard.edu/abs/2010ApJ...710.1032A} {710, 1032}

\bibitem[\protect\citeauthoryear{{Anderson} et~al.,}{{Anderson}
  et~al.}{2008a}]{anderson2008a}
{Anderson} J.,  et~al., 2008a, \mn@doi [\aj] {10.1088/0004-6256/135/6/2055},
  \href {http://adsabs.harvard.edu/abs/2008AJ....135.2055A} {135, 2055}

\bibitem[\protect\citeauthoryear{{Anderson} et~al.,}{{Anderson}
  et~al.}{2008b}]{anderson2008b}
{Anderson} J.,  et~al., 2008b, \mn@doi [\aj] {10.1088/0004-6256/135/6/2114},
  \href {http://adsabs.harvard.edu/abs/2008AJ....135.2114A} {135, 2114}

\bibitem[\protect\citeauthoryear{{Bally}, {O'Dell}  \& {McCaughrean}}{{Bally}
  et~al.}{2000}]{bally2000}
{Bally} J.,  {O'Dell} C.~R.,   {McCaughrean} M.~J.,  2000, \mn@doi [\aj]
  {10.1086/301385}, \href {http://adsabs.harvard.edu/abs/2000AJ....119.2919B}
  {119, 2919}

\bibitem[\protect\citeauthoryear{{Baranov}, {Krasnobaev}  \&
  {Kulikovskii}}{{Baranov} et~al.}{1971}]{baranov1971}
{Baranov} V.~B.,  {Krasnobaev} K.~V.,   {Kulikovskii} A.~G.,  1971, Soviet
  Physics Doklady, \href {http://adsabs.harvard.edu/abs/1971SPhD...15..791B}
  {15, 791}

\bibitem[\protect\citeauthoryear{{Blaauw}}{{Blaauw}}{1961}]{blaauw1961}
{Blaauw} A.,  1961, \bain, \href
  {http://adsabs.harvard.edu/abs/1961BAN....15..265B} {15, 265}

\bibitem[\protect\citeauthoryear{{Bonnell}, {Bate}, {Clarke}  \&
  {Pringle}}{{Bonnell} et~al.}{2001a}]{bonnell2001a}
{Bonnell} I.~A.,  {Bate} M.~R.,  {Clarke} C.~J.,   {Pringle} J.~E.,  2001a,
  \mn@doi [\mnras] {10.1046/j.1365-8711.2001.04270.x}, \href
  {http://adsabs.harvard.edu/abs/2001MNRAS.323..785B} {323, 785}

\bibitem[\protect\citeauthoryear{{Bonnell}, {Clarke}, {Bate}  \&
  {Pringle}}{{Bonnell} et~al.}{2001b}]{bonnell2001b}
{Bonnell} I.~A.,  {Clarke} C.~J.,  {Bate} M.~R.,   {Pringle} J.~E.,  2001b,
  \mn@doi [\mnras] {10.1046/j.1365-8711.2001.04311.x}, \href
  {http://adsabs.harvard.edu/abs/2001MNRAS.324..573B} {324, 573}

\bibitem[\protect\citeauthoryear{{Bonnell}, {Vine}  \& {Bate}}{{Bonnell}
  et~al.}{2004}]{bonnell2004}
{Bonnell} I.~A.,  {Vine} S.~G.,   {Bate} M.~R.,  2004, \mn@doi [\mnras]
  {10.1111/j.1365-2966.2004.07543.x}, \href
  {http://adsabs.harvard.edu/abs/2004MNRAS.349..735B} {349, 735}

\bibitem[\protect\citeauthoryear{{Bovy}}{{Bovy}}{2016}]{bovy2017}
{Bovy} J.,  2016, preprint, \href
  {http://adsabs.harvard.edu/abs/2016arXiv161007610B} {} (\mn@eprint {arXiv}
  {1610.07610})

\bibitem[\protect\citeauthoryear{{Brown} \& {Bomans}}{{Brown} \&
  {Bomans}}{2005}]{brownbomans2005}
{Brown} D.,  {Bomans} D.~J.,  2005, \mn@doi [\aap]
  {10.1051/0004-6361:20041054}, \href
  {http://adsabs.harvard.edu/abs/2005A%26A...439..183B} {439, 183}

\bibitem[\protect\citeauthoryear{{Brownsberger} \& {Romani}}{{Brownsberger} \&
  {Romani}}{2014}]{brownsbergerromani2014}
{Brownsberger} S.,  {Romani} R.~W.,  2014, \mn@doi [\apj]
  {10.1088/0004-637X/784/2/154}, \href
  {http://adsabs.harvard.edu/abs/2014ApJ...784..154B} {784, 154}

\bibitem[\protect\citeauthoryear{{Clark}, {Bonnell}, {Zinnecker}  \&
  {Bate}}{{Clark} et~al.}{2005}]{clark2005}
{Clark} P.~C.,  {Bonnell} I.~A.,  {Zinnecker} H.,   {Bate} M.~R.,  2005,
  \mn@doi [\mnras] {10.1111/j.1365-2966.2005.08942.x}, \href
  {http://adsabs.harvard.edu/abs/2005MNRAS.359..809C} {359, 809}

\bibitem[\protect\citeauthoryear{{Comer{\'o}n} \& {Pasquali}}{{Comer{\'o}n} \&
  {Pasquali}}{2007}]{comeronpasquali2007}
{Comer{\'o}n} F.,  {Pasquali} A.,  2007, \mn@doi [\aap]
  {10.1051/0004-6361:20077304}, \href
  {http://adsabs.harvard.edu/abs/2007A%26A...467L..23C} {467, L23}

\bibitem[\protect\citeauthoryear{{Cox} \& {Lallo}}{{Cox} \&
  {Lallo}}{2012}]{coxlallo2012}
{Cox} C.,  {Lallo} M.,  2012, in Space Telescopes and Instrumentation 2012:
  Optical, Infrared, and Millimeter Wave. p. 844237, \mn@doi{10.1117/12.924900}

\bibitem[\protect\citeauthoryear{{Cox} \& {Niemi}}{{Cox} \&
  {Niemi}}{2011}]{coxniemi2011}
{Cox} C.,  {Niemi} S.-M.,  2011, Technical report, {Evaluation of a
  temperature-based HST focus model}.
Space Telescope Science Institute

\bibitem[\protect\citeauthoryear{{Cruz-Gonz{\'a}lez}, {Recillas-Cruz},
  {Costero}, {Peimbert}  \& {Torres-Peimbert}}{{Cruz-Gonz{\'a}lez}
  et~al.}{1974}]{cruzgonzalez1974}
{Cruz-Gonz{\'a}lez} C.,  {Recillas-Cruz} E.,  {Costero} R.,  {Peimbert} M.,
  {Torres-Peimbert} S.,  1974, \rmxaa, \href
  {http://adsabs.harvard.edu/abs/1974RMxAA...1..211C} {1, 211}

\bibitem[\protect\citeauthoryear{{Damiani} et~al.,}{{Damiani}
  et~al.}{2016}]{damiani2016}
{Damiani} F.,  et~al., 2016, preprint, \href
  {http://adsabs.harvard.edu/abs/2016arXiv160401208D} {} (\mn@eprint {arXiv}
  {1604.01208})

\bibitem[\protect\citeauthoryear{{Davidson} \& {Humphreys}}{{Davidson} \&
  {Humphreys}}{1997}]{davidsonhumphreys1997}
{Davidson} K.,  {Humphreys} R.~M.,  1997, \mn@doi [\araa]
  {10.1146/annurev.astro.35.1.1}, \href
  {http://adsabs.harvard.edu/abs/1997ARA%26A..35....1D} {35, 1}

\bibitem[\protect\citeauthoryear{{Dias}, {Alessi}, {Moitinho}  \&
  {L{\'e}pine}}{{Dias} et~al.}{2002}]{dias2002}
{Dias} W.~S.,  {Alessi} B.~S.,  {Moitinho} A.,   {L{\'e}pine} J.~R.~D.,  2002,
  \mn@doi [\aap] {10.1051/0004-6361:20020668}, \href
  {http://adsabs.harvard.edu/abs/2002A%26A...389..871D} {389, 871}

\bibitem[\protect\citeauthoryear{{Efremov} \& {Elmegreen}}{{Efremov} \&
  {Elmegreen}}{1998}]{efremovelmegreen1998b}
{Efremov} Y.~N.,  {Elmegreen} B.~G.,  1998, \mn@doi [\mnras]
  {10.1046/j.1365-8711.1998.01819.x}, \href
  {http://adsabs.harvard.edu/abs/1998MNRAS.299..588E} {299, 588}

\bibitem[\protect\citeauthoryear{{Elias}, {Alfaro}  \&
  {Cabrera-Ca{\~n}o}}{{Elias} et~al.}{2006}]{elias2006}
{Elias} F.,  {Alfaro} E.~J.,   {Cabrera-Ca{\~n}o} J.,  2006, \mn@doi [\aj]
  {10.1086/505941}, \href {http://adsabs.harvard.edu/abs/2006AJ....132.1052E}
  {132, 1052}

\bibitem[\protect\citeauthoryear{{Feast} \& {Whitelock}}{{Feast} \&
  {Whitelock}}{1997}]{feastwhitelock1997}
{Feast} M.,  {Whitelock} P.,  1997, \mn@doi [\mnras] {10.1093/mnras/291.4.683},
  \href {http://adsabs.harvard.edu/abs/1997MNRAS.291..683F} {291, 683}

\bibitem[\protect\citeauthoryear{{Feigelson} et~al.,}{{Feigelson}
  et~al.}{2011}]{feigelson2011}
{Feigelson} E.~D.,  et~al., 2011, \mn@doi [\apjs] {10.1088/0067-0049/194/1/9},
  \href {http://adsabs.harvard.edu/abs/2011ApJS..194....9F} {194, 9}

\bibitem[\protect\citeauthoryear{{Fitzgerald} \& {Mehta}}{{Fitzgerald} \&
  {Mehta}}{1987}]{fitzgeraldmehta1987}
{Fitzgerald} M.~P.,  {Mehta} S.,  1987, \mn@doi [\mnras]
  {10.1093/mnras/228.3.545}, \href
  {http://adsabs.harvard.edu/abs/1987MNRAS.228..545F} {228, 545}

\bibitem[\protect\citeauthoryear{{France}, {McCandliss}  \& {Lupu}}{{France}
  et~al.}{2007}]{france2007}
{France} K.,  {McCandliss} S.~R.,   {Lupu} R.~E.,  2007, \mn@doi [\apj]
  {10.1086/510481}, \href {http://adsabs.harvard.edu/abs/2007ApJ...655..920F}
  {655, 920}

\bibitem[\protect\citeauthoryear{{Fujii} \& {Portegies Zwart}}{{Fujii} \&
  {Portegies Zwart}}{2011}]{fujiiportegieszwart2011}
{Fujii} M.~S.,  {Portegies Zwart} S.,  2011, \mn@doi [Science]
  {10.1126/science.1211927}, \href
  {http://adsabs.harvard.edu/abs/2011Sci...334.1380F} {334, 1380}

\bibitem[\protect\citeauthoryear{{Gaczkowski}, {Preibisch}, {Ratzka},
  {Roccatagliata}, {Ohlendorf}  \& {Zinnecker}}{{Gaczkowski}
  et~al.}{2013}]{gaczkowski2013}
{Gaczkowski} B.,  {Preibisch} T.,  {Ratzka} T.,  {Roccatagliata} V.,
  {Ohlendorf} H.,   {Zinnecker} H.,  2013, \mn@doi [\aap]
  {10.1051/0004-6361/201219836}, \href
  {http://adsabs.harvard.edu/abs/2013A%26A...549A..67G} {549, A67}

\bibitem[\protect\citeauthoryear{{Gagn{\'e}} et~al.,}{{Gagn{\'e}}
  et~al.}{2011}]{gagne2011}
{Gagn{\'e}} M.,  et~al., 2011, \mn@doi [\apjs] {10.1088/0067-0049/194/1/5},
  \href {http://adsabs.harvard.edu/abs/2011ApJS..194....5G} {194, 5}

\bibitem[\protect\citeauthoryear{{Gaia Collaboration} et~al.,}{{Gaia
  Collaboration} et~al.}{2016a}]{gaia2016a}
{Gaia Collaboration} et~al., 2016a, \mn@doi [\aap]
  {10.1051/0004-6361/201629272}, \href
  {http://adsabs.harvard.edu/abs/2016A%26A...595A...1G} {595, A1}

\bibitem[\protect\citeauthoryear{{Gaia Collaboration} et~al.,}{{Gaia
  Collaboration} et~al.}{2016b}]{gaiabrown2016}
{Gaia Collaboration} et~al., 2016b, \mn@doi [\aap]
  {10.1051/0004-6361/201629512}, \href
  {http://adsabs.harvard.edu/abs/2016A%26A...595A...2G} {595, A2}

\bibitem[\protect\citeauthoryear{{Garc{\'{\i}}a}, {Malaroda}, {Levato},
  {Morrell}  \& {Grosso}}{{Garc{\'{\i}}a} et~al.}{1998}]{garcia1998}
{Garc{\'{\i}}a} B.,  {Malaroda} S.,  {Levato} H.,  {Morrell} N.,   {Grosso} M.,
   1998, \mn@doi [\pasp] {10.1086/316117}, \href
  {http://adsabs.harvard.edu/abs/1998PASP..110...53G} {110, 53}

\bibitem[\protect\citeauthoryear{{G{\'a}sp{\'a}r}, {Su}, {Rieke}, {Balog},
  {Kamp}, {Mart{\'{\i}}nez-Galarza}  \& {Stapelfeldt}}{{G{\'a}sp{\'a}r}
  et~al.}{2008}]{gaspar2008}
{G{\'a}sp{\'a}r} A.,  {Su} K.~Y.~L.,  {Rieke} G.~H.,  {Balog} Z.,  {Kamp} I.,
  {Mart{\'{\i}}nez-Galarza} J.~R.,   {Stapelfeldt} K.,  2008, \mn@doi [\apj]
  {10.1086/523299}, \href {http://adsabs.harvard.edu/abs/2008ApJ...672..974G}
  {672, 974}

\bibitem[\protect\citeauthoryear{{Gies} \& {Bolton}}{{Gies} \&
  {Bolton}}{1986}]{giesbolton1986}
{Gies} D.~R.,  {Bolton} C.~T.,  1986, \mn@doi [\apjs] {10.1086/191118}, \href
  {http://adsabs.harvard.edu/abs/1986ApJS...61..419G} {61, 419}

\bibitem[\protect\citeauthoryear{{Gvaramadze} \& {Bomans}}{{Gvaramadze} \&
  {Bomans}}{2008}]{gvaramadzebomans2008}
{Gvaramadze} V.~V.,  {Bomans} D.~J.,  2008, \mn@doi [\aap]
  {10.1051/0004-6361:200810411}, \href
  {http://adsabs.harvard.edu/abs/2008A%26A...490.1071G} {490, 1071}

\bibitem[\protect\citeauthoryear{{Gvaramadze}, {Kroupa}  \&
  {Pflamm-Altenburg}}{{Gvaramadze} et~al.}{2010}]{gvaramadze2010}
{Gvaramadze} V.~V.,  {Kroupa} P.,   {Pflamm-Altenburg} J.,  2010, \mn@doi
  [\aap] {10.1051/0004-6361/201014871}, \href
  {http://adsabs.harvard.edu/abs/2010A%26A...519A..33G} {519, A33}

\bibitem[\protect\citeauthoryear{{Gvaramadze}, {Pflamm-Altenburg}  \&
  {Kroupa}}{{Gvaramadze} et~al.}{2011a}]{gvaramadze2011a}
{Gvaramadze} V.~V.,  {Pflamm-Altenburg} J.,   {Kroupa} P.,  2011a, \mn@doi
  [\aap] {10.1051/0004-6361/201015656}, \href
  {http://adsabs.harvard.edu/abs/2011A%26A...525A..17G} {525, A17}

\bibitem[\protect\citeauthoryear{{Gvaramadze}, {Kniazev}, {Kroupa}  \&
  {Oh}}{{Gvaramadze} et~al.}{2011b}]{gvaramadze2011b}
{Gvaramadze} V.~V.,  {Kniazev} A.~Y.,  {Kroupa} P.,   {Oh} S.,  2011b, \mn@doi
  [\aap] {10.1051/0004-6361/201117746}, \href
  {http://adsabs.harvard.edu/abs/2011A%26A...535A..29G} {535, A29}

\bibitem[\protect\citeauthoryear{{Gvaramadze}, {Weidner}, {Kroupa}  \&
  {Pflamm-Altenburg}}{{Gvaramadze} et~al.}{2012}]{gvaramadze2012a}
{Gvaramadze} V.~V.,  {Weidner} C.,  {Kroupa} P.,   {Pflamm-Altenburg} J.,
  2012, \mn@doi [\mnras] {10.1111/j.1365-2966.2012.21452.x}, \href
  {http://adsabs.harvard.edu/abs/2012MNRAS.424.3037G} {424, 3037}

\bibitem[\protect\citeauthoryear{{Hills}}{{Hills}}{1980}]{hills1980}
{Hills} J.~G.,  1980, \mn@doi [\apj] {10.1086/157703}, \href
  {http://adsabs.harvard.edu/abs/1980ApJ...235..986H} {235, 986}

\bibitem[\protect\citeauthoryear{{H{\o}g} et~al.,}{{H{\o}g}
  et~al.}{2000}]{hog2000}
{H{\o}g} E.,  et~al., 2000, \aap, \href
  {http://cdsads.u-strasbg.fr/abs/2000A%26A...355L..27H} {355, L27}

\bibitem[\protect\citeauthoryear{{Kaper}, {van Loon}, {Augusteijn},
  {Goudfrooij}, {Patat}, {Waters}  \& {Zijlstra}}{{Kaper}
  et~al.}{1997}]{kaper1997}
{Kaper} L.,  {van Loon} J.~T.,  {Augusteijn} T.,  {Goudfrooij} P.,  {Patat} F.,
   {Waters} L.~B.~F.~M.,   {Zijlstra} A.~A.,  1997, \mn@doi [\apjl]
  {10.1086/310454}, \href {http://adsabs.harvard.edu/abs/1997ApJ...475L..37K}
  {475, L37}

\bibitem[\protect\citeauthoryear{{Kiminki}, {Reiter}  \& {Smith}}{{Kiminki}
  et~al.}{2016}]{kiminki2016}
{Kiminki} M.~M.,  {Reiter} M.,   {Smith} N.,  2016, \mn@doi [\mnras]
  {10.1093/mnras/stw2019}, \href
  {http://adsabs.harvard.edu/abs/2016MNRAS.463..845K} {463, 845}

\bibitem[\protect\citeauthoryear{{Kobulnicky}, {Gilbert}  \&
  {Kiminki}}{{Kobulnicky} et~al.}{2010}]{kobulnicky2010}
{Kobulnicky} H.~A.,  {Gilbert} I.~J.,   {Kiminki} D.~C.,  2010, \mn@doi [\apj]
  {10.1088/0004-637X/710/1/549}, \href
  {http://adsabs.harvard.edu/abs/2010ApJ...710..549K} {710, 549}

\bibitem[\protect\citeauthoryear{{Kobulnicky} et~al.,}{{Kobulnicky}
  et~al.}{2016}]{kobulnicky2016}
{Kobulnicky} H.~A.,  et~al., 2016, preprint, \href
  {http://adsabs.harvard.edu/abs/2016arXiv160902204K} {} (\mn@eprint {arXiv}
  {1609.02204})

\bibitem[\protect\citeauthoryear{{Krist}, {Hook}  \& {Stoehr}}{{Krist}
  et~al.}{2011}]{krist2011}
{Krist} J.~E.,  {Hook} R.~N.,   {Stoehr} F.,  2011, in Optical Modeling and
  Performance Predictions V. p. 81270J, \mn@doi{10.1117/12.892762}

\bibitem[\protect\citeauthoryear{{Krumholz}, {McKee}  \& {Klein}}{{Krumholz}
  et~al.}{2005}]{krumholz2005b}
{Krumholz} M.~R.,  {McKee} C.~F.,   {Klein} R.~I.,  2005, \mn@doi [\nat]
  {10.1038/nature04280}, \href
  {http://adsabs.harvard.edu/abs/2005Natur.438..332K} {438, 332}

\bibitem[\protect\citeauthoryear{{Krumholz}, {Klein}, {McKee}, {Offner}  \&
  {Cunningham}}{{Krumholz} et~al.}{2009}]{krumholz2009}
{Krumholz} M.~R.,  {Klein} R.~I.,  {McKee} C.~F.,  {Offner} S.~S.~R.,
  {Cunningham} A.~J.,  2009, \mn@doi [Science] {10.1126/science.1165857}, \href
  {http://adsabs.harvard.edu/abs/2009Sci...323..754K} {323, 754}

\bibitem[\protect\citeauthoryear{{Lada} \& {Lada}}{{Lada} \&
  {Lada}}{1991}]{ladalada1991}
{Lada} C.~J.,  {Lada} E.~A.,  1991, in {Janes} K.,  ed.,  Astronomical Society
  of the Pacific Conference Series Vol. 13, The Formation and Evolution of Star
  Clusters. pp 3--22

\bibitem[\protect\citeauthoryear{{Lada} \& {Lada}}{{Lada} \&
  {Lada}}{2003}]{ladalada2003}
{Lada} C.~J.,  {Lada} E.~A.,  2003, \mn@doi [\araa]
  {10.1146/annurev.astro.41.011802.094844}, \href
  {http://adsabs.harvard.edu/abs/2003ARA%26A..41...57L} {41, 57}

\bibitem[\protect\citeauthoryear{{Levato}, {Malaroda}, {Garcia}, {Morrell}  \&
  {Solivella}}{{Levato} et~al.}{1990}]{levato1990}
{Levato} H.,  {Malaroda} S.,  {Garcia} B.,  {Morrell} N.,   {Solivella} G.,
  1990, \mn@doi [\apjs] {10.1086/191419}, \href
  {http://adsabs.harvard.edu/abs/1990ApJS...72..323L} {72, 323}

\bibitem[\protect\citeauthoryear{{Lindegren} et~al.,}{{Lindegren}
  et~al.}{2016}]{lindegren2016}
{Lindegren} L.,  et~al., 2016, \mn@doi [\aap] {10.1051/0004-6361/201628714},
  \href {http://adsabs.harvard.edu/abs/2016A%26A...595A...4L} {595, A4}

\bibitem[\protect\citeauthoryear{{Mackey}, {Gvaramadze}, {Mohamed}  \&
  {Langer}}{{Mackey} et~al.}{2015}]{mackey2015}
{Mackey} J.,  {Gvaramadze} V.~V.,  {Mohamed} S.,   {Langer} N.,  2015, \mn@doi
  [\aap] {10.1051/0004-6361/201424716}, \href
  {http://adsabs.harvard.edu/abs/2015A%26A...573A..10M} {573, A10}

\bibitem[\protect\citeauthoryear{{Mackey}, {Haworth}, {Gvaramadze}, {Mohamed},
  {Langer}  \& {Harries}}{{Mackey} et~al.}{2016}]{mackey2016}
{Mackey} J.,  {Haworth} T.~J.,  {Gvaramadze} V.~V.,  {Mohamed} S.,  {Langer}
  N.,   {Harries} T.~J.,  2016, \mn@doi [\aap] {10.1051/0004-6361/201527569},
  \href {http://adsabs.harvard.edu/abs/2016A%26A...586A.114M} {586, A114}

\bibitem[\protect\citeauthoryear{{Mahmud} \& {Anderson}}{{Mahmud} \&
  {Anderson}}{2008}]{mahmudanderson2008}
{Mahmud} N.,  {Anderson} J.,  2008, \mn@doi [\pasp] {10.1086/591290}, \href
  {http://adsabs.harvard.edu/abs/2008PASP..120..907M} {120, 907}

\bibitem[\protect\citeauthoryear{{Massey} \& {Johnson}}{{Massey} \&
  {Johnson}}{1993}]{masseyjohnson1993}
{Massey} P.,  {Johnson} J.,  1993, \mn@doi [\aj] {10.1086/116487}, \href
  {http://adsabs.harvard.edu/abs/1993AJ....105..980M} {105, 980}

\bibitem[\protect\citeauthoryear{{McKee} \& {Tan}}{{McKee} \&
  {Tan}}{2003}]{mckeetan2003}
{McKee} C.~F.,  {Tan} J.~C.,  2003, \mn@doi [\apj] {10.1086/346149}, \href
  {http://adsabs.harvard.edu/abs/2003ApJ...585..850M} {585, 850}

\bibitem[\protect\citeauthoryear{{Moffat} et~al.,}{{Moffat}
  et~al.}{1998}]{moffat1998}
{Moffat} A.~F.~J.,  et~al., 1998, \aap, \href
  {http://adsabs.harvard.edu/abs/1998A%26A...331..949M} {331, 949}

\bibitem[\protect\citeauthoryear{{Moffat} et~al.,}{{Moffat}
  et~al.}{1999}]{moffat1999}
{Moffat} A.~F.~J.,  et~al., 1999, \aap, \href
  {http://adsabs.harvard.edu/abs/1999A%26A...345..321M} {345, 321}

\bibitem[\protect\citeauthoryear{Nelder \& Mead}{Nelder \&
  Mead}{1965}]{neldermead1965}
Nelder J.~A.,  Mead R.,  1965, Computer Journal, 7, 308

\bibitem[\protect\citeauthoryear{{Niemi} \& {Lallo}}{{Niemi} \&
  {Lallo}}{2010}]{niemilallo2010}
{Niemi} S.-M.,  {Lallo} M.,  2010, Technical report, {Phase Retrieval to
  Monitor HST Focus: II. Results Post-Servicing Mission 4}.
Space Telescope Science Institute

\bibitem[\protect\citeauthoryear{{Noriega-Crespo}, {van Buren}  \&
  {Dgani}}{{Noriega-Crespo} et~al.}{1997}]{noriegacrespo1997}
{Noriega-Crespo} A.,  {van Buren} D.,   {Dgani} R.,  1997, \mn@doi [\aj]
  {10.1086/118298}, \href {http://adsabs.harvard.edu/abs/1997AJ....113..780N}
  {113, 780}

\bibitem[\protect\citeauthoryear{{Ochsendorf} \& {Tielens}}{{Ochsendorf} \&
  {Tielens}}{2015}]{ochsendorftielens2015}
{Ochsendorf} B.~B.,  {Tielens} A.~G.~G.~M.,  2015, \mn@doi [\aap]
  {10.1051/0004-6361/201424799}, \href
  {http://adsabs.harvard.edu/abs/2015A%26A...576A...2O} {576, A2}

\bibitem[\protect\citeauthoryear{{Ochsendorf}, {Cox}, {Krijt}, {Salgado},
  {Bern{\'e}}, {Bernard}, {Kaper}  \& {Tielens}}{{Ochsendorf}
  et~al.}{2014a}]{ochsendorf2014a}
{Ochsendorf} B.~B.,  {Cox} N.~L.~J.,  {Krijt} S.,  {Salgado} F.,  {Bern{\'e}}
  O.,  {Bernard} J.~P.,  {Kaper} L.,   {Tielens} A.~G.~G.~M.,  2014a, \mn@doi
  [\aap] {10.1051/0004-6361/201322873}, \href
  {http://adsabs.harvard.edu/abs/2014A%26A...563A..65O} {563, A65}

\bibitem[\protect\citeauthoryear{{Ochsendorf}, {Verdolini}, {Cox}, {Bern{\'e}},
  {Kaper}  \& {Tielens}}{{Ochsendorf} et~al.}{2014b}]{ochsendorf2014b}
{Ochsendorf} B.~B.,  {Verdolini} S.,  {Cox} N.~L.~J.,  {Bern{\'e}} O.,  {Kaper}
  L.,   {Tielens} A.~G.~G.~M.,  2014b, \mn@doi [\aap]
  {10.1051/0004-6361/201423545}, \href
  {http://adsabs.harvard.edu/abs/2014A%26A...566A..75O} {566, A75}

\bibitem[\protect\citeauthoryear{{Ochsendorf}, {Brown}, {Bally}  \&
  {Tielens}}{{Ochsendorf} et~al.}{2015}]{ochsendorf2015}
{Ochsendorf} B.~B.,  {Brown} A.~G.~A.,  {Bally} J.,   {Tielens} A.~G.~G.~M.,
  2015, \mn@doi [\apj] {10.1088/0004-637X/808/2/111}, \href
  {http://adsabs.harvard.edu/abs/2015ApJ...808..111O} {808, 111}

\bibitem[\protect\citeauthoryear{{Oh} \& {Kroupa}}{{Oh} \&
  {Kroupa}}{2016}]{ohkroupa2016}
{Oh} S.,  {Kroupa} P.,  2016, \mn@doi [\aap] {10.1051/0004-6361/201628233},
  \href {http://adsabs.harvard.edu/abs/2016A%26A...590A.107O} {590, A107}

\bibitem[\protect\citeauthoryear{{Patat} \& {Carraro}}{{Patat} \&
  {Carraro}}{2001}]{patatcarraro2001}
{Patat} F.,  {Carraro} G.,  2001, \mn@doi [\mnras]
  {10.1046/j.1365-8711.2001.04576.x}, \href
  {http://adsabs.harvard.edu/abs/2001MNRAS.325.1591P} {325, 1591}

\bibitem[\protect\citeauthoryear{{Penny}, {Gies}, {Hartkopf}, {Mason}  \&
  {Turner}}{{Penny} et~al.}{1993}]{penny1993}
{Penny} L.~R.,  {Gies} D.~R.,  {Hartkopf} W.~I.,  {Mason} B.~D.,   {Turner}
  N.~H.,  1993, \mn@doi [\pasp] {10.1086/133200}, \href
  {http://adsabs.harvard.edu/abs/1993PASP..105..588P} {105, 588}

\bibitem[\protect\citeauthoryear{{Peri}, {Benaglia}, {Brookes}, {Stevens}  \&
  {Isequilla}}{{Peri} et~al.}{2012}]{peri2012}
{Peri} C.~S.,  {Benaglia} P.,  {Brookes} D.~P.,  {Stevens} I.~R.,   {Isequilla}
  N.~L.,  2012, \mn@doi [\aap] {10.1051/0004-6361/201118116}, \href
  {http://adsabs.harvard.edu/abs/2012A%26A...538A.108P} {538, A108}

\bibitem[\protect\citeauthoryear{{Peri}, {Benaglia}  \& {Isequilla}}{{Peri}
  et~al.}{2015}]{peri2015}
{Peri} C.~S.,  {Benaglia} P.,   {Isequilla} N.~L.,  2015, \mn@doi [\aap]
  {10.1051/0004-6361/201424676}, \href
  {http://adsabs.harvard.edu/abs/2015A%26A...578A..45P} {578, A45}

\bibitem[\protect\citeauthoryear{{Pflamm-Altenburg} \&
  {Kroupa}}{{Pflamm-Altenburg} \& {Kroupa}}{2010}]{pflammaltenburgkroupa2010}
{Pflamm-Altenburg} J.,  {Kroupa} P.,  2010, \mn@doi [\mnras]
  {10.1111/j.1365-2966.2010.16376.x}, \href
  {http://adsabs.harvard.edu/abs/2010MNRAS.404.1564P} {404, 1564}

\bibitem[\protect\citeauthoryear{{Poveda}, {Ruiz}  \& {Allen}}{{Poveda}
  et~al.}{1967}]{poveda1967}
{Poveda} A.,  {Ruiz} J.,   {Allen} C.,  1967, Boletin de los Observatorios
  Tonantzintla y Tacubaya, \href
  {http://adsabs.harvard.edu/abs/1967BOTT....4...86P} {4, 86}

\bibitem[\protect\citeauthoryear{{Povich}, {Benjamin}, {Whitney}, {Babler},
  {Indebetouw}, {Meade}  \& {Churchwell}}{{Povich} et~al.}{2008}]{povich2008}
{Povich} M.~S.,  {Benjamin} R.~A.,  {Whitney} B.~A.,  {Babler} B.~L.,
  {Indebetouw} R.,  {Meade} M.~R.,   {Churchwell} E.,  2008, \mn@doi [\apj]
  {10.1086/592565}, \href {http://adsabs.harvard.edu/abs/2008ApJ...689..242P}
  {689, 242}

\bibitem[\protect\citeauthoryear{{Povich} et~al.,}{{Povich}
  et~al.}{2011}]{povich2011b}
{Povich} M.~S.,  et~al., 2011, \mn@doi [\apjs] {10.1088/0067-0049/194/1/6},
  \href {http://adsabs.harvard.edu/abs/2011ApJS..194....6P} {194, 6}

\bibitem[\protect\citeauthoryear{{Preibisch} et~al.,}{{Preibisch}
  et~al.}{2011}]{preibisch2011c}
{Preibisch} T.,  et~al., 2011, \mn@doi [\aap] {10.1051/0004-6361/201116781},
  \href {http://adsabs.harvard.edu/abs/2011A%26A...530A..34P} {530, A34}

\bibitem[\protect\citeauthoryear{Press, Teukolsky, Vetterling  \&
  Flannery}{Press et~al.}{1992}]{press1992}
Press W.~H.,  Teukolsky S.~A.,  Vetterling W.~T.,   Flannery B.~P.,  1992,
  Numerical Recipes in C (2nd Ed.): The Art of Scientific Computing.
Cambridge University Press, New York, NY, USA

\bibitem[\protect\citeauthoryear{{Rebolledo} et~al.,}{{Rebolledo}
  et~al.}{2016}]{rebolledo2016}
{Rebolledo} D.,  et~al., 2016, \mn@doi [\mnras] {10.1093/mnras/stv2776}, \href
  {http://adsabs.harvard.edu/abs/2016MNRAS.456.2406R} {456, 2406}

\bibitem[\protect\citeauthoryear{{Reed}}{{Reed}}{2003}]{reed2003}
{Reed} B.~C.,  2003, \mn@doi [\aj] {10.1086/374771}, \href
  {http://adsabs.harvard.edu/abs/2003AJ....125.2531R} {125, 2531}

\bibitem[\protect\citeauthoryear{{Reiter}, {Smith}, {Kiminki}, {Bally}  \&
  {Anderson}}{{Reiter} et~al.}{2015a}]{reiter2015a}
{Reiter} M.,  {Smith} N.,  {Kiminki} M.~M.,  {Bally} J.,   {Anderson} J.,
  2015a, \mn@doi [\mnras] {10.1093/mnras/stv177}, \href
  {http://adsabs.harvard.edu/abs/2015MNRAS.448.3429R} {448, 3429}

\bibitem[\protect\citeauthoryear{{Reiter}, {Smith}, {Kiminki}  \&
  {Bally}}{{Reiter} et~al.}{2015b}]{reiter2015b}
{Reiter} M.,  {Smith} N.,  {Kiminki} M.~M.,   {Bally} J.,  2015b, \mn@doi
  [\mnras] {10.1093/mnras/stv634}, \href
  {http://adsabs.harvard.edu/abs/2015MNRAS.450..564R} {450, 564}

\bibitem[\protect\citeauthoryear{{Schilbach} \& {R{\"o}ser}}{{Schilbach} \&
  {R{\"o}ser}}{2008}]{schilbachroser2008}
{Schilbach} E.,  {R{\"o}ser} S.,  2008, \mn@doi [\aap]
  {10.1051/0004-6361:200809936}, \href
  {http://adsabs.harvard.edu/abs/2008A%26A...489..105S} {489, 105}

\bibitem[\protect\citeauthoryear{{Sch{\"o}nrich}, {Binney}  \&
  {Dehnen}}{{Sch{\"o}nrich} et~al.}{2010}]{schonrich2010}
{Sch{\"o}nrich} R.,  {Binney} J.,   {Dehnen} W.,  2010, \mn@doi [\mnras]
  {10.1111/j.1365-2966.2010.16253.x}, \href
  {http://adsabs.harvard.edu/abs/2010MNRAS.403.1829S} {403, 1829}

\bibitem[\protect\citeauthoryear{{Sexton}, {Povich}, {Smith}, {Babler}, {Meade}
   \& {Rudolph}}{{Sexton} et~al.}{2015}]{sexton2015}
{Sexton} R.~O.,  {Povich} M.~S.,  {Smith} N.,  {Babler} B.~L.,  {Meade} M.~R.,
   {Rudolph} A.~L.,  2015, \mn@doi [\mnras] {10.1093/mnras/stu2143}, \href
  {http://adsabs.harvard.edu/abs/2015MNRAS.446.1047S} {446, 1047}

\bibitem[\protect\citeauthoryear{{Smith}}{{Smith}}{2006a}]{smith2006a}
{Smith} N.,  2006a, \mn@doi [\mnras] {10.1111/j.1365-2966.2006.10007.x}, \href
  {http://adsabs.harvard.edu/abs/2006MNRAS.367..763S} {367, 763}

\bibitem[\protect\citeauthoryear{{Smith}}{{Smith}}{2006b}]{smith2006b}
{Smith} N.,  2006b, \mn@doi [\apj] {10.1086/503766}, \href
  {http://adsabs.harvard.edu/abs/2006ApJ...644.1151S} {644, 1151}

\bibitem[\protect\citeauthoryear{{Smith} \& {Brooks}}{{Smith} \&
  {Brooks}}{2008}]{smithbrooks2008}
{Smith} N.,  {Brooks} K.~J.,  2008, in {Reipurth} B.,  ed., , Handbook of Star
  Forming Regions, Volume II.
ASP, San Francisco, CA, p.~138

\bibitem[\protect\citeauthoryear{{Smith}, {Egan}, {Carey}, {Price}, {Morse}  \&
  {Price}}{{Smith} et~al.}{2000}]{smith2000}
{Smith} N.,  {Egan} M.~P.,  {Carey} S.,  {Price} S.~D.,  {Morse} J.~A.,
  {Price} P.~A.,  2000, \mn@doi [\apjl] {10.1086/312578}, \href
  {http://adsabs.harvard.edu/abs/2000ApJ...532L.145S} {532, L145}

\bibitem[\protect\citeauthoryear{{Smith}, {Bally}  \& {Walborn}}{{Smith}
  et~al.}{2010a}]{smith2010a}
{Smith} N.,  {Bally} J.,   {Walborn} N.~R.,  2010a, \mn@doi [\mnras]
  {10.1111/j.1365-2966.2010.16520.x}, \href
  {http://adsabs.harvard.edu/abs/2010MNRAS.405.1153S} {405, 1153}

\bibitem[\protect\citeauthoryear{{Smith} et~al.,}{{Smith}
  et~al.}{2010b}]{smith2010b}
{Smith} N.,  et~al., 2010b, \mn@doi [\mnras]
  {10.1111/j.1365-2966.2010.16792.x}, \href
  {http://adsabs.harvard.edu/abs/2010MNRAS.406..952S} {406, 952}

\bibitem[\protect\citeauthoryear{{Sohn}, {Anderson}  \& {van der Marel}}{{Sohn}
  et~al.}{2012}]{sohn2012}
{Sohn} S.~T.,  {Anderson} J.,   {van der Marel} R.~P.,  2012, \mn@doi [\apj]
  {10.1088/0004-637X/753/1/7}, \href
  {http://adsabs.harvard.edu/abs/2012ApJ...753....7S} {753, 7}

\bibitem[\protect\citeauthoryear{{Sota}, {Ma{\'{\i}}z Apell{\'a}niz},
  {Morrell}, {Barb{\'a}}, {Walborn}, {Gamen}, {Arias}  \& {Alfaro}}{{Sota}
  et~al.}{2014}]{sota2014}
{Sota} A.,  {Ma{\'{\i}}z Apell{\'a}niz} J.,  {Morrell} N.~I.,  {Barb{\'a}}
  R.~H.,  {Walborn} N.~R.,  {Gamen} R.~C.,  {Arias} J.~I.,   {Alfaro} E.~J.,
  2014, \mn@doi [\apjs] {10.1088/0067-0049/211/1/10}, \href
  {http://adsabs.harvard.edu/abs/2014ApJS..211...10S} {211, 10}

\bibitem[\protect\citeauthoryear{{Stone}}{{Stone}}{1991}]{stone1991}
{Stone} R.~C.,  1991, \mn@doi [\aj] {10.1086/115880}, \href
  {http://adsabs.harvard.edu/abs/1991AJ....102..333S} {102, 333}

\bibitem[\protect\citeauthoryear{{Tetzlaff}, {Neuh{\"a}user}  \&
  {Hohle}}{{Tetzlaff} et~al.}{2011}]{tetzlaff2011}
{Tetzlaff} N.,  {Neuh{\"a}user} R.,   {Hohle} M.~M.,  2011, \mn@doi [\mnras]
  {10.1111/j.1365-2966.2010.17434.x}, \href
  {http://adsabs.harvard.edu/abs/2011MNRAS.410..190T} {410, 190}

\bibitem[\protect\citeauthoryear{{Tutukov}}{{Tutukov}}{1978}]{tutukov1978}
{Tutukov} A.~V.,  1978, \aap, \href
  {http://adsabs.harvard.edu/abs/1978A%26A....70...57T} {70, 57}

\bibitem[\protect\citeauthoryear{{Vijapurkar} \& {Drilling}}{{Vijapurkar} \&
  {Drilling}}{1993}]{vijapurkardrilling1993}
{Vijapurkar} J.,  {Drilling} J.~S.,  1993, \mn@doi [\apjs] {10.1086/191849},
  \href {http://cdsads.u-strasbg.fr/abs/1993ApJS...89..293V} {89, 293}

\bibitem[\protect\citeauthoryear{{Walborn}}{{Walborn}}{1995}]{walborn1995}
{Walborn} N.~R.,  1995, in {Niemela} V.,  {Morrell} N.,   {Feinstein} A.,  eds,
   Revista Mexicana de Astronomia y Astrofisica Conference Series Vol. 2,
  Revista Mexicana de Astronomia y Astrofisica Conference Series. p.~51

\bibitem[\protect\citeauthoryear{{Walborn} \& {Hesser}}{{Walborn} \&
  {Hesser}}{1975}]{walbornhesser1975}
{Walborn} N.~R.,  {Hesser} J.~E.,  1975, \mn@doi [\apj] {10.1086/153720}, \href
  {http://adsabs.harvard.edu/abs/1975ApJ...199..535W} {199, 535}

\bibitem[\protect\citeauthoryear{{Walborn} et~al.,}{{Walborn}
  et~al.}{2002a}]{walborn2002a}
{Walborn} N.~R.,  et~al., 2002a, \mn@doi [\aj] {10.1086/339831}, \href
  {http://adsabs.harvard.edu/abs/2002AJ....123.2754W} {123, 2754}

\bibitem[\protect\citeauthoryear{{Walborn}, {Danks}, {Vieira}  \&
  {Landsman}}{{Walborn} et~al.}{2002b}]{walborn2002b}
{Walborn} N.~R.,  {Danks} A.~C.,  {Vieira} G.,   {Landsman} W.~B.,  2002b,
  \mn@doi [\apjs] {10.1086/339373}, \href
  {http://adsabs.harvard.edu/abs/2002ApJS..140..407W} {140, 407}

\bibitem[\protect\citeauthoryear{{Walborn}, {Smith}, {Howarth}, {Vieira Kober},
  {Gull}  \& {Morse}}{{Walborn} et~al.}{2007}]{walborn2007}
{Walborn} N.~R.,  {Smith} N.,  {Howarth} I.~D.,  {Vieira Kober} G.,  {Gull}
  T.~R.,   {Morse} J.~A.,  2007, \mn@doi [\pasp] {10.1086/511756}, \href
  {http://adsabs.harvard.edu/abs/2007PASP..119..156W} {119, 156}

\bibitem[\protect\citeauthoryear{{Wilkin}}{{Wilkin}}{2000}]{wilkin2000}
{Wilkin} F.~P.,  2000, \mn@doi [\apj] {10.1086/308576}, \href
  {http://adsabs.harvard.edu/abs/2000ApJ...532..400W} {532, 400}

\bibitem[\protect\citeauthoryear{{Winston}, {Wolk}, {Bourke}, {Megeath},
  {Gutermuth}  \& {Spitzbart}}{{Winston} et~al.}{2012}]{winston2012}
{Winston} E.,  {Wolk} S.~J.,  {Bourke} T.~L.,  {Megeath} S.~T.,  {Gutermuth}
  R.,   {Spitzbart} B.,  2012, \mn@doi [\apj] {10.1088/0004-637X/744/2/126},
  \href {http://adsabs.harvard.edu/abs/2012ApJ...744..126W} {744, 126}

\bibitem[\protect\citeauthoryear{{Wright}, {Parker}, {Goodwin}  \&
  {Drake}}{{Wright} et~al.}{2014}]{wright2014}
{Wright} N.~J.,  {Parker} R.~J.,  {Goodwin} S.~P.,   {Drake} J.~J.,  2014,
  \mn@doi [\mnras] {10.1093/mnras/stt2232}, \href
  {http://adsabs.harvard.edu/abs/2014MNRAS.438..639W} {438, 639}

\bibitem[\protect\citeauthoryear{{Wright}, {Bouy}, {Drew}, {Sarro}, {Bertin},
  {Cuillandre}  \& {Barrado}}{{Wright} et~al.}{2016}]{wright2016}
{Wright} N.~J.,  {Bouy} H.,  {Drew} J.~E.,  {Sarro} L.~M.,  {Bertin} E.,
  {Cuillandre} J.-C.,   {Barrado} D.,  2016, \mn@doi [\mnras]
  {10.1093/mnras/stw1148}, \href
  {http://adsabs.harvard.edu/abs/2016MNRAS.460.2593W} {460, 2593}

\bibitem[\protect\citeauthoryear{{Wu}, {Zhou}, {Ma}  \& {Du}}{{Wu}
  et~al.}{2009}]{wu2009}
{Wu} Z.-Y.,  {Zhou} X.,  {Ma} J.,   {Du} C.-H.,  2009, \mn@doi [\mnras]
  {10.1111/j.1365-2966.2009.15416.x}, \href
  {http://adsabs.harvard.edu/abs/2009MNRAS.399.2146W} {399, 2146}

\bibitem[\protect\citeauthoryear{{Zacharias}, {Finch}, {Girard}, {Henden},
  {Bartlett}, {Monet}  \& {Zacharias}}{{Zacharias}
  et~al.}{2013}]{zacharias2013}
{Zacharias} N.,  {Finch} C.~T.,  {Girard} T.~M.,  {Henden} A.,  {Bartlett}
  J.~L.,  {Monet} D.~G.,   {Zacharias} M.~I.,  2013, \mn@doi [\aj]
  {10.1088/0004-6256/145/2/44}, \href
  {http://cdsads.u-strasbg.fr/abs/2013AJ....145...44Z} {145, 44}

\bibitem[\protect\citeauthoryear{{Zinnecker}}{{Zinnecker}}{1982}]{zinnecker1982}
{Zinnecker} H.,  1982, \mn@doi [Annals of the New York Academy of Sciences]
  {10.1111/j.1749-6632.1982.tb43399.x}, \href
  {http://adsabs.harvard.edu/abs/1982NYASA.395..226Z} {395, 226}

\bibitem[\protect\citeauthoryear{{de Wit}, {Testi}, {Palla}, {Vanzi}  \&
  {Zinnecker}}{{de Wit} et~al.}{2004}]{dewit2004}
{de Wit} W.~J.,  {Testi} L.,  {Palla} F.,  {Vanzi} L.,   {Zinnecker} H.,  2004,
  \mn@doi [\aap] {10.1051/0004-6361:20040454}, \href
  {http://adsabs.harvard.edu/abs/2004A%26A...425..937D} {425, 937}

\bibitem[\protect\citeauthoryear{{de Wit}, {Testi}, {Palla}  \&
  {Zinnecker}}{{de Wit} et~al.}{2005}]{dewit2005}
{de Wit} W.~J.,  {Testi} L.,  {Palla} F.,   {Zinnecker} H.,  2005, \mn@doi
  [\aap] {10.1051/0004-6361:20042489}, \href
  {http://adsabs.harvard.edu/abs/2005A%26A...437..247D} {437, 247}

\bibitem[\protect\citeauthoryear{{di Nino}, {Makidon}, {Lallo}, {Sahu},
  {Sirianni}  \& {Casertano}}{{di Nino} et~al.}{2008}]{dinino2008}
{di Nino} D.,  {Makidon} R.~B.,  {Lallo} M.,  {Sahu} K.~C.,  {Sirianni} M.,
  {Casertano} S.,  2008, Technical report, {HST Focus Variations with
  Temperature}.
Space Telescope Science Institute

\bibitem[\protect\citeauthoryear{{van Buren} \& {McCray}}{{van Buren} \&
  {McCray}}{1988}]{vanburenmccray1988}
{van Buren} D.,  {McCray} R.,  1988, \mn@doi [\apjl] {10.1086/185184}, \href
  {http://adsabs.harvard.edu/abs/1988ApJ...329L..93V} {329, L93}

\bibitem[\protect\citeauthoryear{{van Buren}, {Noriega-Crespo}  \&
  {Dgani}}{{van Buren} et~al.}{1995}]{vanburen1995}
{van Buren} D.,  {Noriega-Crespo} A.,   {Dgani} R.,  1995, \mn@doi [\aj]
  {10.1086/117739}, \href {http://adsabs.harvard.edu/abs/1995AJ....110.2914V}
  {110, 2914}

\makeatother
\end{thebibliography}

%%%%%%%%%%%%%%%%%%%%%%%%%%%%%%%%%%%%%%%%%%%%%%%%%%%%%%%%%%%%%%%%%%%%%%%

\bsp	
\label{lastpage}
\end{document}